\documentclass[a4paper,11pt]{article}
\pdfoutput=1 

\usepackage{jheppub} 
\usepackage{bm,amssymb,slashed,graphicx,multirow,soul,mathtools,xspace,array,tikz,amsmath,siunitx}
\usepackage[compat=1.1.0]{tikz-feynman}   
\usepackage{float}                   
\allowdisplaybreaks
\usepackage{ bbold }
\usepackage{caption,subcaption}
\usepackage{braket}

\definecolor{nicered}{rgb}{0.7,0.1,0.1}
\definecolor{nicegreen}{rgb}{0.1,0.5,0.1}
\definecolor{violet}{rgb}{0.7,0.3,0.3}
\hypersetup{colorlinks,citecolor= nicegreen,linkcolor= nicered}

\newcommand{\todo}[1]{{\color{red} \ifmmode\else[todo]\fi #1}}

\newcommand{\lp}{\left(}
\newcommand{\rp}{\right)}

\newcommand{\Xm}{X_{\rm max}}

\newcommand{\beq}{\begin{equation} }
\newcommand{\eeq}{\end{equation}} 
\newcommand{\bi}{\begin{itemize} }
\newcommand{\ei}{\end{itemize} }
\newcommand{\gcm}{~{\rm g}/{\rm cm}^2} 

\definecolor{Red}{rgb}{1.,0.,0.}
\definecolor{Grn}{rgb}{0.,0.75,0.}
\definecolor{Blu}{rgb}{0.,0.,1.}
\definecolor{Pink}{rgb}{1,0.08,0.58}

\usepackage[T1]{fontenc} 

\usepackage{amsmath,amssymb,epsfig,ulem,color,slashed}
\allowdisplaybreaks  

\title{\boldmath Learning the Composition of Ultra High Energy Cosmic Rays}

\author[1,2]{Bla\v{z} Bortolato,}
\author[1,2]{Jernej F. Kamenik,}
\author[1]{Michele Tammaro,}

\affiliation[1]{Jo\v{z}ef Stefan Institute, Jamova 39, 1000 Ljubljana, Slovenia}
\affiliation[2]{Faculty of Mathematics and Physics, University of Ljubljana, Jadranska 19, 1000 Ljubljana, Slovenia}

\emailAdd{blaz.bortolato@ijs.si}
\emailAdd{jernej.kamenik@cern.ch}
\emailAdd{michele.tammaro@ijs.si}

\abstract{
We apply statistical inference on the Pierre Auger Open Data to discern for the first time the full mass composition of cosmic rays at different energies. Working with longitudinal electromagnetic profiles of cosmic ray showers, in particular their peaking depths $\Xm$,  we employ central moments of the $\Xm$ distributions as features to discriminate between different shower compositions. We find that already the first few moments entail the most relevant information to infer the primary cosmic ray mass spectrum. Our approach, based on an unbinned likelihood, allows us to consistently account for sources of statistical uncertainties due to finite datasets, both measured and simulated, as well as systematic effects. Finally, we provide a quantitative comparison of different high energy hadronic interaction models available in the atmospheric shower simulation codes.
}

\begin{document} 

\maketitle

\flushbottom

\newpage

\section{Introduction}
\label{sec:intro}

Ultra High Energy Cosmic Rays (UHECRs) are nucleons and ionized nuclei colliding with the Earth's atmosphere at energies $E \gtrsim 10^{18}$ eV. Although the first observation of an UHECR dates back to 1963 \cite{PhysRevLett.10.146}, there are still many open questions on the topic (see Refs.~\cite{Mollerach:2017idb,Anchordoqui:2018qom} for recent reviews). Firstly, the spatial distribution of UHECR sources is poorly known. The reconstruction of a UHECR direction is particularly cumbersome since the galactic and extra-galactic magnetic fields deflect these particles during propagation. Secondly, the mechanisms that accelerate them to such high energies have not yet been identified. Possibilities range from acceleration due to supermassive black holes or SuperNovae explosions to galaxy collisions~\cite{Torres:2004hk}. Thirdly, there are large uncertainties regarding the mass composition of UHECRs, that is which kind of particles constitute UHECRs at different energies.

Direct observation of UHECRs by balloon or spacecraft experiments becomes highly inefficient due to their steep energy spectrum: at $E\sim10^{13}$ eV the flux of incoming cosmic ray is $\phi\sim10^{3}~{\rm km}^{-2}~{\rm s}^{-1}$, while it drops to $\phi\sim10^{-2}~{\rm km}^{-2}~{\rm yr}^{-1}$ at $E\sim10^{20}$ eV \cite{Zyla:2020zbs,PierreAuger:2020qqz}. Thus one must rely on ground based detectors to observe the biproducts of UHECRs interacting with the atmosphere. The energetic particle, referred as the primary, scatters with nuclei in the higher layers of the Earth's atmosphere and produces a cascade of secondary particles, which carry a fraction of the primary energy and propagate onward, scattering again or decaying into tertiary particles, and so on. This cascade is called an Extensive Air Shower (EAS).

Currently, the largest operating EAS observatory is the Pierre Auger Observatory~\cite{PierreAuger:2015eyc}, which is composed of 1660 water Cherenkov detectors, called Surface Detectors (SD), at 1.5 km distance from each other, covering an area of 3000 km$^2$ in the Pampa desert of Argentina, and 4 Fluorescent Detectors (FD). It recently completed its 13 years-long first data taking run and is upgrading its detectors for a planned Run 2~\cite{Stasielak:2021hjm}.

The complete evolution of an EAS is a complicated process involving EM and hadronic processes across many energy scales. In this work we focus the observed {\it longitudinal profile} of an EAS, that is the intensity of fluorescent light emitted by nuclei in the atmosphere, typically nytrogen, excited by the passage of charged particles, and measured as a function of the s.c. slant depth ($X$) of the shower. In general, the longitudinal profile has a clear peak, at $\Xm$, corresponding to the point of maximum population of $e^\pm$ in the shower evolution. It depends strongly on the energy and species of the primary. In particular, assuming that the energy $E$ of the primary particle, with atomic number $A$, is shared equally by all the nucleons, it can be shown that $\langle \Xm \rangle \propto \ln E - \ln A$~\cite{Mollerach:2017idb}.

Despite the simple relation, in practice the identification of primaries is not straightforward. Since we cannot observe nucleus - nucleus scattering at ultra-high energies directly, simulations of EAS development have to rely on extrapolations from lower energy measurements using models of hadronic interactions. This leads to significant systematic uncertainties as evidenced by discrepancies between predictions based on different hadronic models. For example, the EPOS-LHC model~\cite{Pierog:2013ria}, based on extrapolations of scattering cross sections from LHC data, gives significantly different results for $\langle \Xm \rangle$ than the QGSJet~\cite{Ostapchenko:2007qb, Ostapchenko:2010vb} model, which instead uses a phenomenological approach to describe the non-perturbative parton cascades. In addition, the actual EAS development is influenced by many fluctuating parameters, such as the first interaction height and incidence angle, or the varying atmospheric conditions along the shower depth. Finally, the longitudinal profiles can only be detected at sufficiently low levels of environmental photon background, such as in moonless nights, thus limiting the available statistics. 

A recent analysis of Pierre Auger Observatory data, in particular of the energy dependence of the average peak position $\langle \Xm \rangle$, suggests that the mass spectrum of UHECR is dominated by protons at energies $E\lesssim10^{18}$ EeV, while it tends towards heavier nuclei at higher energies~\cite{PierreAuger:2019phh} (page 86). However, existing methods do not allow to infer the complete primary composition from the available data. 
In Ref.~\cite{Lipari:2020uca} the spectrum (in particular the binned distributions of $\Xm$ within fixed energy bins) was fitted to a limited mixture of primaries, with the best fit primary fractions depending on the energy and on the hadronic model used in simulations. Using a mixture of five elements, (p, He, N, Si, Fe), high energy Auger data is best accommodated by a combination of Si and Fe initiated showers. However, as shown in Refs.~\cite{Arsene:2020ago,Arsene:2021inm}, the presence of intermediate elements, such as Ne or C, can affect the results, and including up to 8 elements improved the overall goodness of fit. While the choice of mixtures restricted to a few possible elements is quite arbitrary, these studies seem to confirm indications that the high energy tail of the UHECR spectrum cannot be explained by exclusively light primaries.

In present work we improve and extend these previous studies in several ways. 
Our goal is to infer the complete composition that best describes the measured $\Xm$ distribution at different energies and systematically investigate the uncertainties due to simulation and modelling limitations. We define the composition as
\beq
w = \lp w_p, w_{He},~\dots,~w_{Fe} \rp\,, \qquad \sum_P w_P = 1\,,
\eeq
where the primary index $P$ scans over all the 26 possible primaries, from the hydrogen nucleus ($A=1,~P=p$) to iron  ($A=56,~P=Fe$). The composition $w$ is then a 26 dimensional vector of weights. 
Given the low statistics available (especially when working with the Auger Open Data \cite{the_pierre_auger_collaboration_2021_4487613}) and the complexity of the problem, instead of working with binned $\Xm$ distributions directly, we characterize each distribution by its first few central moments: the mean $\langle \Xm \rangle$, the standard deviation $\sigma_{\Xm}$, the skew $\gamma_{\Xm}$, etc. This approach has several advantages: Firstly, it avoids issues of binning sparse distributions as we can compute the moments directly for the unbinned $\Xm$ distributions. In addition it allows us to systematically incorporate additional qualitative features of the $\Xm$ distributions in terms of the moments expansion. We explore their increasing discriminative power, both in resolving the primary composition as well as in comparing predictions of different hadronic models. Finally, we are able to transparently incorporate effects of systematic uncertainties, such as finite simulation samples, on the inferred compositions. 

The problem of estimating the primary composition from data is one of statistical inference: the most probable composition $w^*$ in an energy bin and for a given hadronic model is the one that maximizes the likelihood of reproducing the (moments of) Auger $\Xm$ data with a $w$ weighted mixture of simulated showers, where likelihood maximization is performed on the parameter $w$, and all systematic uncertainties are treated via nuisance parameters. 
Working with a full 26 component weight vector $w$ implies finding the maximum likelihood and the relevant confidence regions on a 26 dimensional manifold. We employ several methods primarily developed for applications in machine learning, such as stochastic minimization techniques together with nested sampling algorithms (see e.g.~\cite{doi:10.1063/1.1835238}) and bootstrapping~\cite{EfroTibs93}, to tackle this otherwise computationally prohibitively demanding task.

The manuscript is organized as follows: In Sec.~\ref{sec:DataAndSim} we give an overview of the Auger Open Data set used and the Monte Carlo simulations of EAS. We introduce the decomposition of $\Xm$ distributions into central moments, discuss their properties and uncertainties, and compare predictions within different hadronic models, in Sec.~\ref{sec:SetUp}. In Sec.~\ref{sec:Composition} we construct our primary composition likelihood model and describe the computational methods which allow us to solve it. Our main results on the inferred UHECR compositions in different energy bins and for different hadronic models are presented in Sec.~\ref{sec:Composition:Results}. Finally, in Sec.~\ref{sec:Conclusions} we summarize our main conclusions and explore possible future directions. As we use illustrative examples in the main text for our discussions, we collect all the relevant additional plots and figures in Appendix~\ref{app:extraplots}.

\section{Data and Simulations}
\label{sec:DataAndSim}

\subsection{Pierre Auger 2021 Open Data}
\label{subsec:OpenData}

\begin{figure}
\begin{center}
        \includegraphics[width=0.6\linewidth]{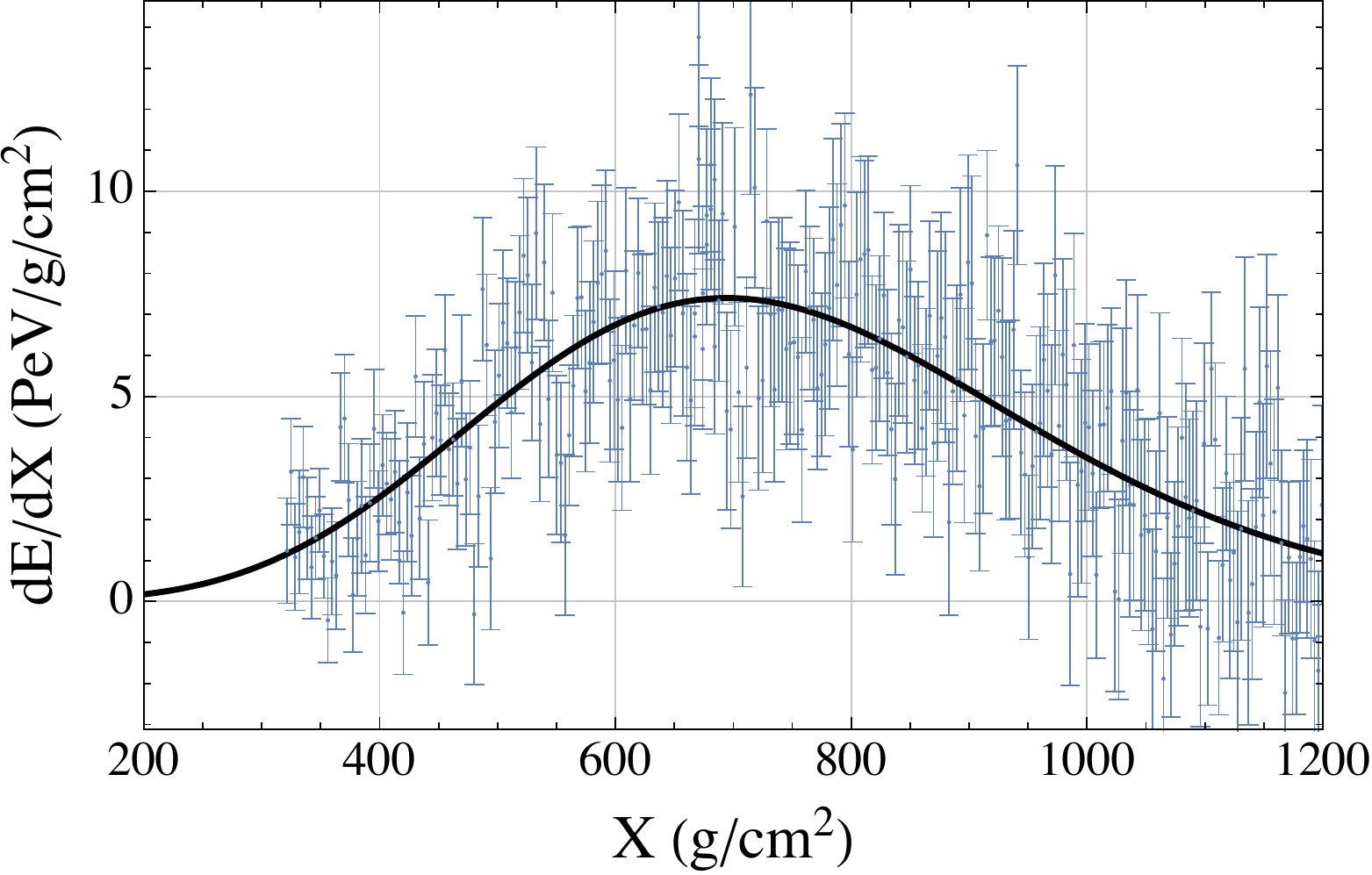}
        \caption{Deposited energy per slant depth. The blue dots represent the FD measurements with uncertainties, while the black line is the fitted GH function.}
        \label{fig:data_summary}
\end{center}
\end{figure}

The Pierre Auger 2021 Open Data \cite{the_pierre_auger_collaboration_2021_4487613} consists of 22731 SD measurements of EAS, which we refer to as Non-Hybrid (NH) showers, and of 3156 ``brass hybrid'' (BH) events, that is showers that have been recorded simultaneously by the SD and the FD. Of these BH, 1602 are called ``golden hybrids'' (GH), with independent SD and Hybrid reconstructions. This dataset amounts to 10\% of the total data collected by the Pierre Auger Collaboration and has already been subject to high-quality selection criteria and cuts, as it is used by the Collaboration itself for their data analysis. Details of the data selection can be found for example in Ref.~\cite{PierreAuger:2020qqz}. Here we review the properties of FD measurements and $\Xm$ distributions and their fitting functions. The former is shown in Fig.~\ref{fig:data_summary} for a sample shower in the Open Data, ${\rm id} = 112636786700$, while the latter is shown in comparison with selected simulations in Fig.~\ref{fig:comparison}, see next section for more details on the simulations.

The electromagnetic signal observed by FDs is strictly related to the primary composition of the UHECR. The energy deposited in the FD is measured as function of the air mass traversed by the shower, the slant depth $X$. This profile can be described by the Gaisser-Hillas parametrization \cite{1977ICRC....8..353G}
\beq\label{eq:FDfit:GH}
f_{GH}(X) = \lp \frac{{\rm d}E}{{\rm d}X} \rp_{\text{max}} \lp \frac{X - X_0}{X_{\text{max}} - X_0} \rp^{\frac{X_{\text{max}} - X_0}{\lambda}} \exp\lp \frac{X_{\text{max}} - X}{\lambda} \rp\,,
\eeq
where $\lp {\rm d}E/{\rm d}X \rp_{\text{max}}$ is the maximum energy deposit at the corresponding depth $X_{\text{max}}$, while $X_0$ and $\lambda$ are two fit parameters. This profile is universal and does not depend on the primary particle \cite{Andringa:2012ju}, however its parameters contain information on the mass composition. Indeed, it can be shown that $X_{\text{max}}$ is proportional to the logarithm of the primary atomic mass number $A$~\cite{Mollerach:2017idb}. On the other hand, the exact shape of the $X_{\text{max}}$ distribution is strongly affected by the intrinsic fluctuations on the first primary scattering in the atmosphere and by the uncertainties on the proton-air cross section at ultra-high energies \cite{PierreAuger:2014sui,PierreAuger:2018gfc}. Nevertheless, the $X_{\text{max}}$ represents the most reliable observable to infer the composition of UHECR. 

The longitudinal profile is usually studied in terms of the shifted and normalized distribution $f_{GH}'(X')$: the depth is shifted as $X' = X - X_{\text{max}}$ such that every curve is centered at zero, and the total distribution is normalized by the energy deposit at the maximum, $\lp {\rm d}E/{\rm d}X \rp_{\text{max}}$. Introducing also the parameters $L = \sqrt{|X_0'|\lambda}$ and $R = \sqrt{\lambda/|X_0'|}$, with $X_0' = X_0 - X_{\text{max}}$~\cite{PierreAuger:2018gfc}, we have 
\beq
f_{GH}'(X') = \lp 1 + R \frac{X'}{L}  \rp^{R^{-2}} \exp \lp - \frac{X'}{LR} \rp\,.
\eeq
The latter distribution is similar to a Gaussian centered at zero and with a standard deviation $L$, but distorted by a multiplicative term governed by $R$. In Fig.~\ref{fig:data_summary} we show the measured longitudinal profile for the sample shower, together with the (unshifted) GH fit as a black line. For each hybrid shower, both the FD data and the parameter set $\{ X_{\text{max}},\lp {\rm d}E/{\rm d}X \rp_{\text{max}},L,R \}$, with their respective uncertainties, are provided in the Open Data.

In particular, we are interested in the distribution of the set $\{ \Xm^1,\dots,\Xm^N \}$ in a fixed energy bin. Assuming that each point is normally distributed around the mean value, $\Xm^j$, with width given by the uncertainty, $\delta\Xm^j$, where $j=1,\dots,N$, we build the PDF
\beq\label{eq:PDF:Auger}
P_{{\rm Aug}}(\Xm) = \frac1N \sum_{j=1}^N {\cal N}\lp \Xm ~|~ \Xm^j,\delta\Xm^j \rp\,.
\eeq

\subsection{Shower Simulations}
\label{subsec:Simulation}

We use CORSIKA 7.7401 \cite{1998cmcc.book.....H} to simulate EAS from UHECR and their longitudinal profiles. The latter is then fitted to the Gaisser-Hillas function, Eq.~\eqref{eq:FDfit:GH}, to extract the depth of the maximum, $\Xm$. Since it has been shown in Ref.~\cite{PierreAuger:2014sui} that $\Xm$ is independent from the incidence angle of the UHECR, we only simulate EAS for incident UHECRs perpendicular to the atmosphere.

For a selected set of inputs $S$, the result of $N$ simulations is a distribution of values $\{ \Xm^{\rm sim} \}_N(S)$. Among the many tunable parameters of the CORSIKA code, here we restrict the discussion to three main inputs: the primary nucleus $Z$, the energy range $E$ and the hadronic model $H$. 
The primary particles simulated are nuclei with proton number $Z$ ranging from $Z=1$ (proton) to $Z=26$ (iron). Since CORSIKA takes as input both the proton number $Z$ and the mass number $A$, we consider for each element only the $A$ for the most abundant stable isotope as to avoid ambiguities. 
The Auger Observatory has observed showers with primary energies up to $\sim 10^{20}$ EeV. However, to both avoid excessive use of computational resources and have a reasonable set of data available from the public release, we restrict our study to primary energies $E \leq 5$ EeV. We additionally divide this set into three bins, namely $E \leq 1$ EeV, $1 < E \leq 2$ EeV and $2 < E \leq 5$ EeV. In the Open Data, these three intervals contain 1002 (345), 1233 (696) and 653 (421) BH (GH) showers respectively. Within each bin, we simulate showers using a flat distribution in energy.
Finally, we consider four available hadronic models in CORSIKA: QGSJET01~\cite{Kalmykov:1993qe}, QGSJetII-04~\cite{Ostapchenko:2010vb}, EPOS~\cite{Pierog:2013ria} and Sibyll 2.3c~\cite{Riehn:2017mfm}. The choice of the hadronic model substantially affects the results of the simulations, as different treatments of the hadronic interactions at very high energies affect the proton-air cross section and the evolution of the hadronic component of the shower. In turn, these (model-dependent) calculations predict distributions of $X_{\text{max}}$. 

In total, we perform 2000 simulations per primary, energy bin and hadronic model, for a total of 624000 simulated showers. We use the parametrization in Eq.~\eqref{eq:FDfit:GH} to extract the value of $\Xm^{\rm sim}$ from each simulated longitudinal shower profile. Although the uncertainty from the fit procedure is quite small, we take this into account and denote it as $\delta\Xm^{\rm sim}$.

\begin{center}
\begin{figure}[!t]
        \includegraphics[width=1.\linewidth]{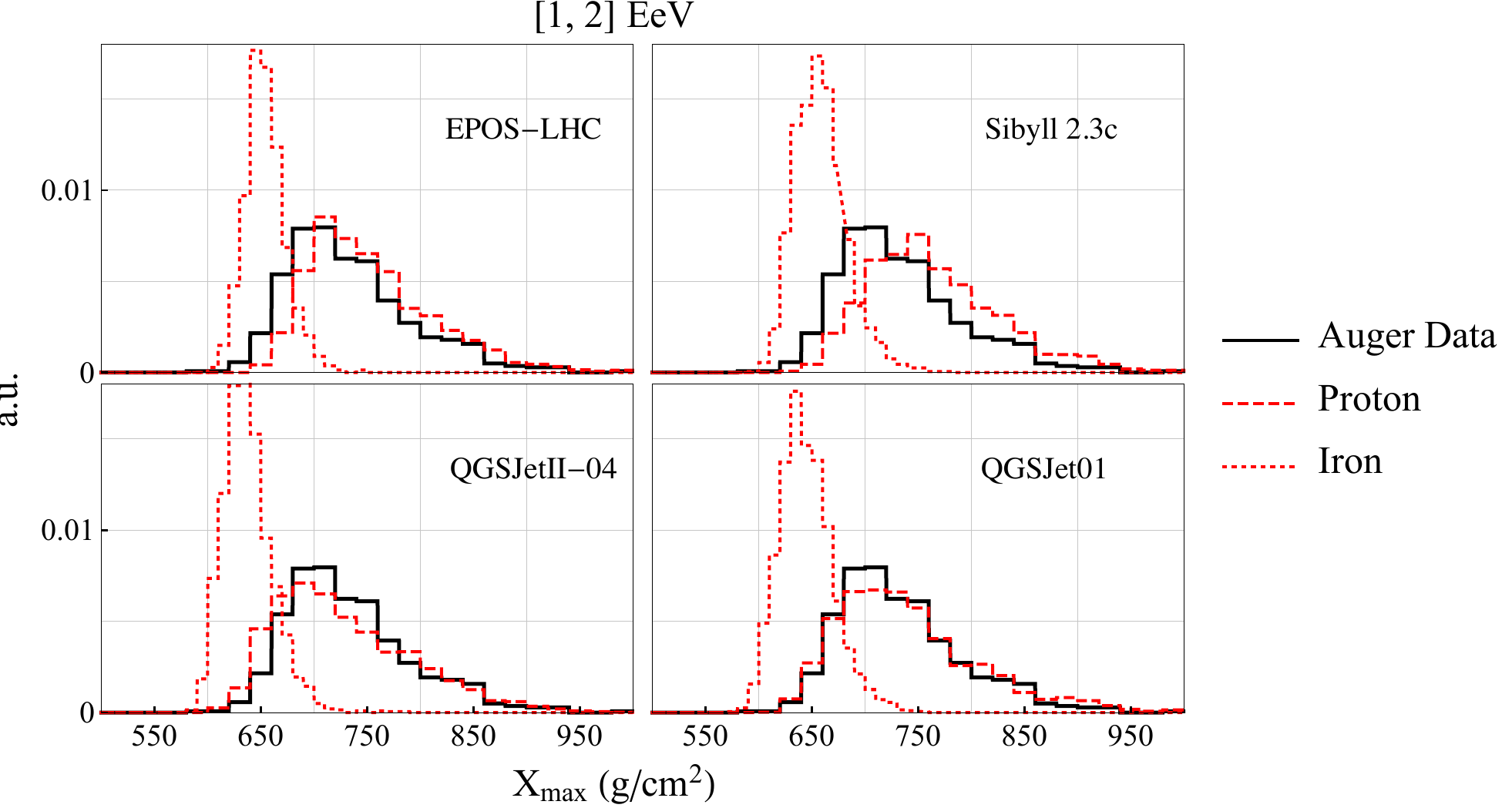}
        \caption{Comparison of $X_{\text{max}}$ distributions for real data (black line) and simulated data, with proton (red dashed) and iron (red dotted) as primary, for energies in the interval $[1,2]$ EeV. 
        }
        \label{fig:comparison}
\end{figure}
\end{center}

In Fig.~\ref{fig:comparison}, we compare the (binned) probability distribution function (PDF) of $X_{\text{max}}$ for the GH showers (black line) in the energy interval $[1,2]$~EeV to simulated showers with proton (red dashed line) and iron (red dotted line) as primaries. We observe that in addition to shifts in the peaks of the distributions between proton and iron, the simulations consistently predict narrower distributions of $\Xm$ for iron ($\sigma_{\Xm}\sim10-20\gcm$), compared to the proton distributions ($\sigma_{\Xm}\sim40-90\gcm$), however the difference varies considerably between simulations based on different hadronic models. 
We thus conduct our analysis with all four models separately and perform a quantitative and systematic study of differences between hadronic models in Sec.~\ref{subsec:ModelComparison}.

\subsection{Detector Effects}
\label{subsec:DetectorEffects}

From each set of $\Xm$ simulated in a fixed energy bin, we can build the respective PDF, as already done in Eq.~\eqref{eq:PDF:Auger}, by summing the single normal distributions. However, in order to be compared to the measured data, the simulation outputs need to be convoluted with the experimental detector acceptance ($\epsilon$), and resolution ($R$). These effects also constitute the main contribution to the total systematic uncertainty. Constructing the PDF as in Eq.~\eqref{eq:PDF:Auger} thus naturally includes these errors in any computation involving $P(X)$.

The inclusion of detector effects reshapes the distribution of simulated $\Xm$, and acceptance in particular also changes its normalization. For a fixed set of inputs $S = \{ Z, E, H\}$, the corresponding PDF is given by
\beq\label{eq:PDF:simulation:nonN}
P_{{\rm sim}} (\Xm ~|~ S) = \frac{1}{\tilde N} \sum_j \int {\rm d}\tilde X ~ {\cal N}\lp \tilde X ~|~ \Xm^j, \delta\Xm^j \rp\times R(\Xm - \tilde X)\times\epsilon(\tilde X)\,,
\eeq
where the index $j=1,\dots,N$ scans over the simulated showers and $\tilde N$ is a normalizing constant. The acceptance $\epsilon$ is parameterized as a piece-wise function of $X$, with a central constant part, and two exponential extremes
\beq 
\epsilon(X) = 
\begin{cases}
\exp\lp \frac{X - x_1}{\lambda_1} \rp & X\leq x_1 \\
1 & x_1\leq X\leq x_2 \\
\exp\lp -\frac{X - x_2}{\lambda_2} \rp & X\leq x_2
\end{cases}
\eeq
while the resolution $R$ is parameterized as a combination of two normal distributions of $X - \tilde X$, centered around the origin,
\beq 
R\lp X\rp = R_3 ~ {\cal N}\lp X  ~|~ 0, R_1 \rp + (1 - R_3) ~ {\cal N}\lp X  ~|~ 0, R_2 \rp\,,
\eeq
where the sets of parameters $(x_1,\lambda_1,x_2,\lambda_2)$ and $(R_1, R_2, R_3)$ depend on the energy bin.
The full detailed description of detector effects and other sources of systematic errors, together with  the numerical values of $\epsilon$ and $R$ parameters, are given in Ref.~\cite{PierreAuger:2014sui}. 

The integral in Eq.~\eqref{eq:PDF:simulation:nonN} depends solely on the specific shower considered and the energy bin where it falls. For later convenience we define the quantity
\beq\label{eq:Fnotation}
F_j(\Xm~|~ S) \equiv \int {\rm d}\tilde X ~ {\cal N}\lp \tilde X ~|~ \Xm^j,\delta \Xm^j \rp\times R(\Xm - \tilde X)\times\epsilon(\tilde X)\,,
\eeq
and rewrite the PDF as
\beq\label{eq:PDF:simulation:nonN:Fnotation}
P_{{\rm sim}} (\Xm ~|~ S) = \frac{1}{\tilde N} \sum_j F_j(\Xm~|~ S)\,.
\eeq
The advantage of this notation is twofold: firstly, we have written the PDF as a sum of single integrals, which only need to be evaluated once for each $S$; secondly, the systematic uncertainty from detector effects is included in each $F_j$ in a transparent way. The latter is studied in more detail in Section~\ref{subsec:Bootstrapping}.

Note that the Auger Collaboration divides the energy in smaller bins, namely in the intervals $\log_{10}E \in[e,~e+0.1]$, where $e = 17.8,~17.9,\dots,20$, and the energy in measured in eV. The numerical values of the efficiency and smearing functions are given in Ref.~\cite{PierreAuger:2014sui} for each of these bins. We take them into account in building Eq.~\eqref{eq:PDF:simulation:nonN}, then combine the results in the larger energy bins we defined in Sec.~\ref{subsec:Simulation}. As an example, the first two energy bins in Auger, $B_1\equiv\log_{10}E\in[17.8,~17.9]$ and $B_2\equiv\log_{10}E\in[17.9,~18.0]$, are contained in our first energy bin, $E_1\equiv E\in[0.6,~1]$ EeV. We can then build $P_{{\rm sim}}(X~|~E_1)$ (we omit the other inputs here) by subdividing the summation over the index $j$, into a sum over $j_1$ and $j_2$, where each index scans over the $N_1$ and $N_2$ showers in the bins $B_1$ and $B_2$ respectively. Namely we have 
\beq 
P_{{\rm sim}}(\Xm~|~E_1) = \frac{1}{N_1+N_2}\Big( N_1 ~ P_{{\rm sim}}(\Xm~|~B_1) + N_2 ~P_{{\rm sim}}(\Xm~|~B_2)  \Big)\,.
\eeq

\section{Central Moments of $\Xm$ Distributions}
\label{sec:SetUp}

\subsection{Moment Decomposition}
\label{subsec:MomentDecomposition}

Given a data set $\{X_{\rm max}\}$ containing $\Xm^i$ with $i = 1,\dots,N$, its mean and central moments are defined as
\begin{align}
z_1 &\equiv \langle X_{\rm max} \rangle = \frac1N \sum_{i=1}^N X_{{\rm max},i}\,, \label{eq:Moments:Mean} \\
z_n &= \frac1N \sum_{i=1}^N \lp X_{{\rm max},i} - z_1 \rp^n  \label{eq:Moments:Higher} \,,
\end{align}
where $\{ \Xm \}$ can be either the observed data set, $\{ \Xm^{\rm Aug} \}$, or a simulated set with fixed inputs, $\{ \Xm^{\rm sim} \}$.
The central moments can be used to characterize a distribution. In particular, higher central moments describe the spread and shape about its mean. Effectively we build a map
\begin{equation}\label{eq:DimRedMap}
    \mathcal{G}: P(X) \rightarrow 
    \begin{pmatrix}
    z_1\\
    \vdots\\
    z_n\\
    \end{pmatrix}.
\end{equation}
which reduces the dimension of each $X_{\rm max}$ set from $N$ numbers, where typically $N\sim10^3$, to a set of $n$ moments. 
While dimensionality reduction can be achieved in several ways, for example by binning or training a neural network, this particular map reduction exhibits excellent performance, offers transparent interpretation, and is suitable also for low statistics samples. In particular, in Section~\ref{subsec:MomentCorrelations} we argue that $n=3$ already entails the most relevant distribution information; furthermore, we show in Sec.~\ref{subsec:Comparison:4primaries} that 3 moments are sufficient to reproduce existing results in the literature obtained by considering the full (binned) $X_{\rm max}$ distribution.

Once we fix the energy $E$ and the hadronic model $H$ in our simulations, we obtain a PDF for each primary $Z$, $P(\Xm~|~Z)$, see Eq.~\eqref{eq:PDF:simulation:nonN}.
We can write the $n$-th ordinary moment of this distribution as
\beq\label{eq:Moment:fromPDF}
\langle \Xm^n \rangle_Z = \frac{\int P(\Xm~|~Z)~\Xm^n~{\rm d}\Xm}{\int P(\Xm~|~Z)~{\rm d}\Xm} \,.
\eeq
The above expression can be further simplified. We can formally define the ``$n$-th moment'' for each individual shower $j$ as
\beq\label{eq:calF}
{\cal F}_j^n(Z) = \int F_j(\Xm|Z)~\Xm^n~{\rm d}\Xm\,,
\eeq
where we have used the expression in Eq.~\eqref{eq:PDF:simulation:nonN:Fnotation}. Defining 
\beq
\frac1N\sum_j {\cal F}_j^0(Z) \equiv \Delta_Z = \int P(\Xm~|~Z)~{\rm d}\Xm \,,
\eeq
leads to the final expression
\beq 
\langle \Xm^n \rangle_Z =  \frac{\frac1N\sum_j{\cal F}_j^n(Z)}{\Delta_Z}\,.
\eeq
This form is highly convenient for numerical evaluation, as the integrals ${\cal F}_j^n$ only need to be computed once for a given dataset.
$\Delta_Z$ represents the renormalization of the $\Xm$ distribution due to detector effects. In particular detector efficiency and smearing in general lead to $\Delta_Z < 1$. Numerically we find this effect to be at the $\mathcal O (1\%)$ level, that is $\Delta_Z \gtrsim 0.99$, and thus negligible with our statistics. Nonetheless, we keep this notation in the remainder of the paper as this is in general the proper normalization factor for the evaluation of moments.

The $n$-th central moment $z_n$ can then be  written as a linear combination of $\langle \Xm^{m\leq n} \rangle$. The first moment is the mean, $z_1 \equiv \langle \Xm \rangle$, while for $n>1$ we can write in general
\beq 
z_n = \langle (\Xm - \langle \Xm \rangle)^n \rangle = \sum^n_{k = 0}\binom{n}{k} \langle \Xm^{n-k} \rangle  (-1)^k \langle \Xm \rangle^k\,.
\eeq
Explicitly, for the first four moments we have
\beq
\label{eq:CentralMoments}
\begin{split}
    z_{1} &= \langle \Xm \rangle,\\
    z_{2} &= \langle \Xm^2 \rangle -  \langle \Xm \rangle^2,\\
    z_{3} &= \langle \Xm^3 \rangle - 3 \langle \Xm^2 \rangle \langle \Xm \rangle + 2 \langle \Xm \rangle^3,\\
    z_{4} &= \langle \Xm^4 \rangle - 4 \langle \Xm^3 \rangle \langle \Xm \rangle
    +6 \langle \Xm^2 \rangle \langle \Xm \rangle^2 - 3 \langle \Xm \rangle^4\,,
\end{split}
\eeq
where the dependence on $Z$ has been omitted. Finally, for a composition $w$, the total $n$-th moment can be written as
\beq
\label{eq:Wmoments}
\langle \Xm^n \rangle(w) = \dfrac{\sum_{Z} \langle \Xm^n \rangle_Z ~ \Delta_Z ~ w_Z}{\sum_{Z} \Delta_Z ~ w_Z}\,.
\eeq
Note that while an ordinary moment $\langle \Xm^n \rangle$ of a composition is a weighted average of the same moments $\langle \Xm^n \rangle_Z$ of individual components, this is in general not true for central moments, which have to be computed through Eq.~\eqref{eq:CentralMoments}.

\subsection{Correlations}
\label{subsec:MomentCorrelations}

We perform the decomposition of the $\Xm$ distributions into moments with the aim of capturing their most discriminating features when inferring the UHECR composition. 
However, if moments $z_n$ and $z_{n+1}$ are highly correlated for a fixed set of simulation inputs ($Z,E,H$), then $z_{n+1}$ actually does not provide much additional information on the distribution shape with respect to $z_n$. In order to have a quantitative measure of such correlations, we perform linear fits between two sets of moments, $\{ z_n \}$ and $\{ z_{n+1} \}$, for a fixed energy bin, primary CR, and hadronic model, obtained via the bootstrapping procedure described in Section~\ref{subsec:Bootstrapping}. 
For each moment pair we compute the correlation coefficient, $R$. If $R\to1$, the two moments are highly linearly correlated, while $R\to0$ indicates weak to no correlation.

\begin{center}
\begin{figure}[!t]
        \includegraphics[width=.95\linewidth]{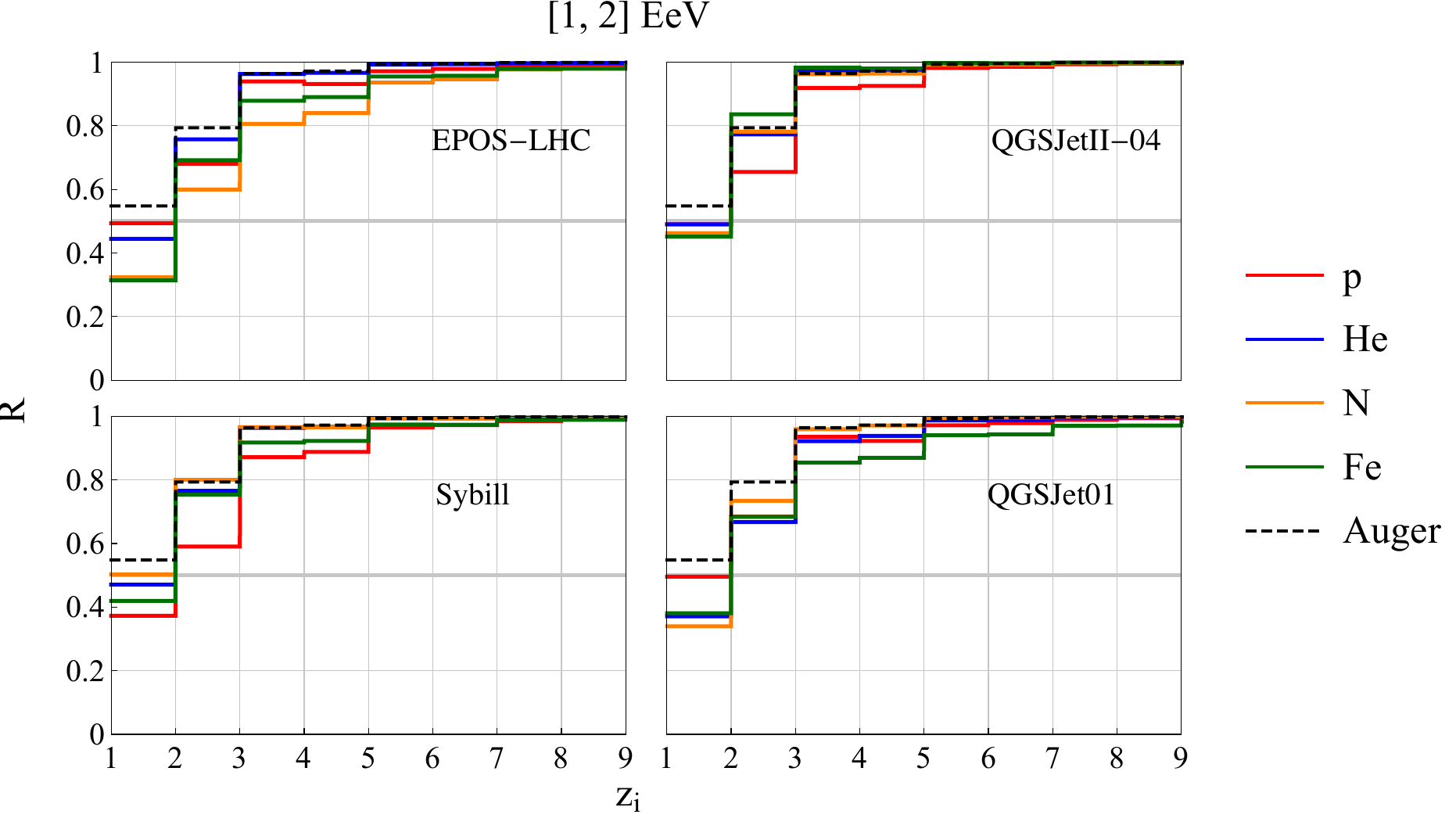}
        \caption{Correlation coefficients between consecutive moments for four simulated primaries, proton (p, red), helium (He, blue), nitrogen (N, orange) and iron (Fe, dark green), and for Auger data (black dashed). The thick horizontal grey line indicates $R = 0.5$. Each segment shows the value of $R$ for $z_i$ and $z_{i+1}$; e.g. the first segment in each plot shows $R$ for the fit of $z_1$ and $z_2$, the second segment for the fit of $z_2$ and $z_3$ and so on. }
        \label{fig:MomentsCorrelation}
\end{figure}
\end{center}

In Fig.~\ref{fig:MomentsCorrelation} we show the correlation coefficients of nine consecutive moments for four different primaries, p, He, N and Fe, simulated with each hadronic model in the $[1,2]$ EeV energy bin. Each segment indicates the value of $R$ for the linear fit between $z_i$ and $z_{i+1}$. From the plots we can see that even $z_1$ and $z_2$ are not completely uncorrelated, with $R\sim 0.5$. Nonetheless, the second moment is expected to provide significant complementary information to the mean ($\langle \Xm \rangle$).  The third moment shows further increased correlation with the second one, as indicated by the values  in the range $R \sim (0.6-0.8)$, depending on the primary and hadronic model. Going beyond the third moment in the expansion, we find strong correlations between $z_i$ and $z_{i+1}$, that is $R\sim1$.\footnote{We have checked explicitly that the conclusion remains qualitatively the same also for s.c. normalized (dimensionless) moments defined as $z_i/z_2^{(i/2)}$ for $i>2$. 
} These higher moments are thus expected to add increasingly marginal additional information to the analysis. We test this hypothesis explicitly in Section~\ref{sec:Composition:Results} by comparing inferred compositions and their uncertainties at different truncations of the $z_i$ expansion.

\subsection{Model Comparison}
\label{subsec:ModelComparison}

One important feature of parameterizing $\Xm$ distributions in terms of their moments is that it allows for a systematic and transparent comparison of different high energy hadronic interaction models discussed in Section~\ref{subsec:Simulation}. 
To illustrate this, we fix the primary energy $E$, and compare sets of $z_n$ computed from simulated showers using different combinations of the hadronic model $H$ and primary $Z$ (again, these $\{z_n(H,Z)\}$ sets are obtained using the bootstrapping procedure described in Section~\ref{subsec:Bootstrapping}). 
To quantify the comparison, we estimate the s.c. Hellinger distance defined between pairs of PDFs. Given two probability density functions $p_1(x)$, $p_2(x)$, the Hellinger distance $\mathcal H$ is defined via
\beq\label{eq:HellingerDist}
\mathcal H^2(p_1,p_2) = \int \lp \sqrt{p_1(x)} - \sqrt{p_2(x)} \rp^2 {\rm d}x = 1 - \int \sqrt{p_1(x)p_2(x)} {\rm d}x\,.
\eeq
Thus, $\mathcal H$ defines a metric in the space of PDFs, bounded in the range $[0,1]$. Intuitively, one can think of two PDFs being ``distant'' when $\mathcal H^2(p_1,p_2)\to1$, as $p_1$ would assign zero probability to all points $x$ where $p_2(x)>0$ and vice versa. Conversely the two PDFs are ``near'' when $\mathcal H^2(p_1,p_2)\to0$. 
When both PDFs are gaussians, $p_1(x)\sim{\cal N}(\mu_1,\sigma_1)$ and $p_2(x)\sim{\cal N}(\mu_2,\sigma_2)$, the integral in \eqref{eq:HellingerDist} can be solved analytically yielding
\beq
\mathcal H^2(p_1,p_2) = 1 - \sqrt{\frac{2\sigma_1\sigma_2}{\sigma_1^2 + \sigma_2^2}}\exp\left[-\frac14 \frac{(\mu_1 - \mu_2)^2}{\sigma_1^2 + \sigma_2^2} \right]\,.
\label{eq:Hellinger}
\eeq

\begin{center}
\begin{figure}[!t]
        ~~~~~~~~~~~~\includegraphics[width=0.8\linewidth]{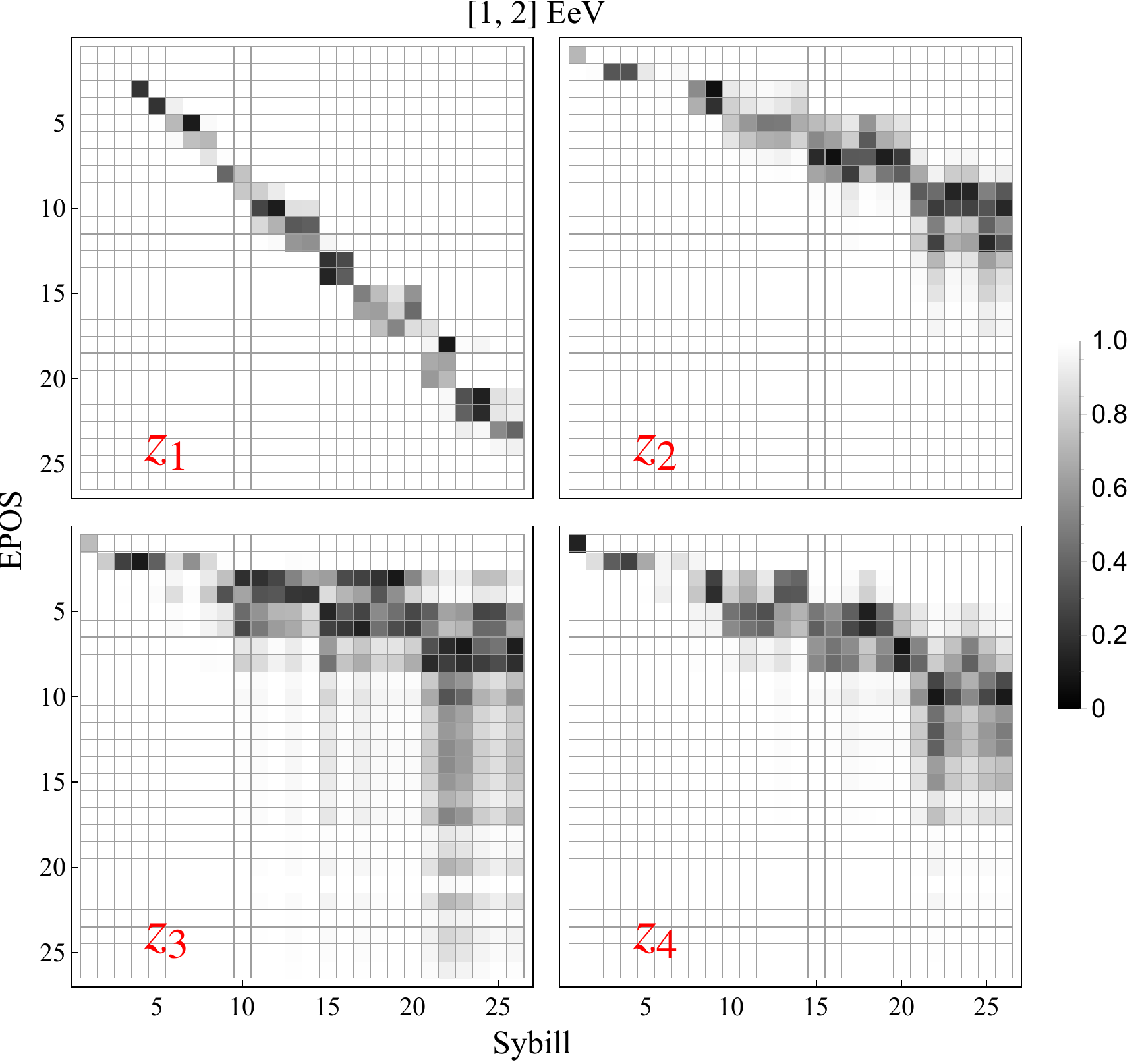}
        \caption{Hellinger distance $\mathcal H^{ab}_{ij}(z_n)$ between EPOS ($a$) and Sybill ($b$) models and different primaries $i,j$ (both axis represent primary atomic numbers) for the first four central moments $z_n$ of $\Xm$ distributions of simulated UHECRs with energies within $[1,2]$~EeV. See text for details.}
        \label{fig:EPOSvsSIB}
\end{figure}
\end{center}

For each set $\{z_n(H_{},Z_{})\}$, we thus first approximate (fit) the relevant PDFs as gaussians $p(z_n)) \sim \mathcal N(\mu(z_n),\sigma(z_n))$. 
Then for pairs $p(z_n(H_a,Z_i))$ $p(z_n(H_b,Z_j))$, the Hellinger distance $\mathcal H^{ab}_{ij}(z_n) \equiv \sqrt{\mathcal H^2 (p(z_n(H_a,Z_i)),p(z_n(H_b,Z_j)))}$ for each moment $z_n$ as given in Eq.~\eqref{eq:Hellinger} quantifies systematic differences between simulations based upon the two hadronic models. In particular $\mathcal H^{ab}_{i=j}(z_n) \gg \mathcal H^{ab}_{i\neq j}(z_n)$ (for some $i\neq j$) would directly indicate a systematic relative bias between a pair of models related to primary UHECR inference based on the $z_n$ moment. As an example, in Fig.~\ref{fig:EPOSvsSIB} we show the matrix of Hellinger distances for the first four central moments of EPOS and Sibyll, simulated in the $[1,2]$~EeV energy bin.\footnote{Similar results comparing all hadronic model pairs in the same energy bin, are presented in Appendix~\ref{app:extraplots}.} 
 We observe that while the two hadronic models agree relatively well for the first moment (the smallest values of $\mathcal H$ are close to the diagonal), already at $z_2$ their predictions start to systematically exhibit significant relative bias (smallest values of $\mathcal H$ are shifted systematically away from the diagonal) as well as spread (more cells with $\mathcal H \ll 1$).
Both effects become even more pronounced for $z_3$ and $z_4$. Relative bias implies that the two hadronic models could in principle infer different UHECR compositions for the same measured $z_{2,3,4}$. The effect is however partially off-set by the spread, which signifies that the resolution or discriminating power between primaries diminishes, i.e. distributions of $z_{2,3,4}$ for different (neighbouring) primaries (even for the same hadronic model, see Appendix~\ref{app:extraplots}) are more and more similar. What is important however is that at least for very distant primaries (with very different atomic numbers $Z$) higher moments retain significant discriminatory power ($\mathcal H \sim 1$).

\section{Inferring Composition of UHECRs}
\label{sec:Composition}

\subsection{Evaluation of Uncertainties via Bootstrapping}
\label{subsec:Bootstrapping}

The total uncertainty of simulated $\{\Xm \}$ PDFs and subsequently of $z_i$ receives two main contributions: the systematic error from detector effects, as described in Sec.~\ref{subsec:DetectorEffects}, and the statistical uncertainty due to the finite number of showers available. While the former is included in our definition of the PDF, see Eq.~\eqref{eq:PDF:simulation:nonN}, we need to evaluate the latter in a consistent way. Our strategy consists of bootstrapping the PDFs of simulated showers and evaluating the resulting widths of moment distributions. In the following discussion we assume that the energy bin and hadronic model are fixed and keep only the primary particle $Z$ as variable.

\begin{figure}[t]
\centering
        \includegraphics[width=.7\linewidth]{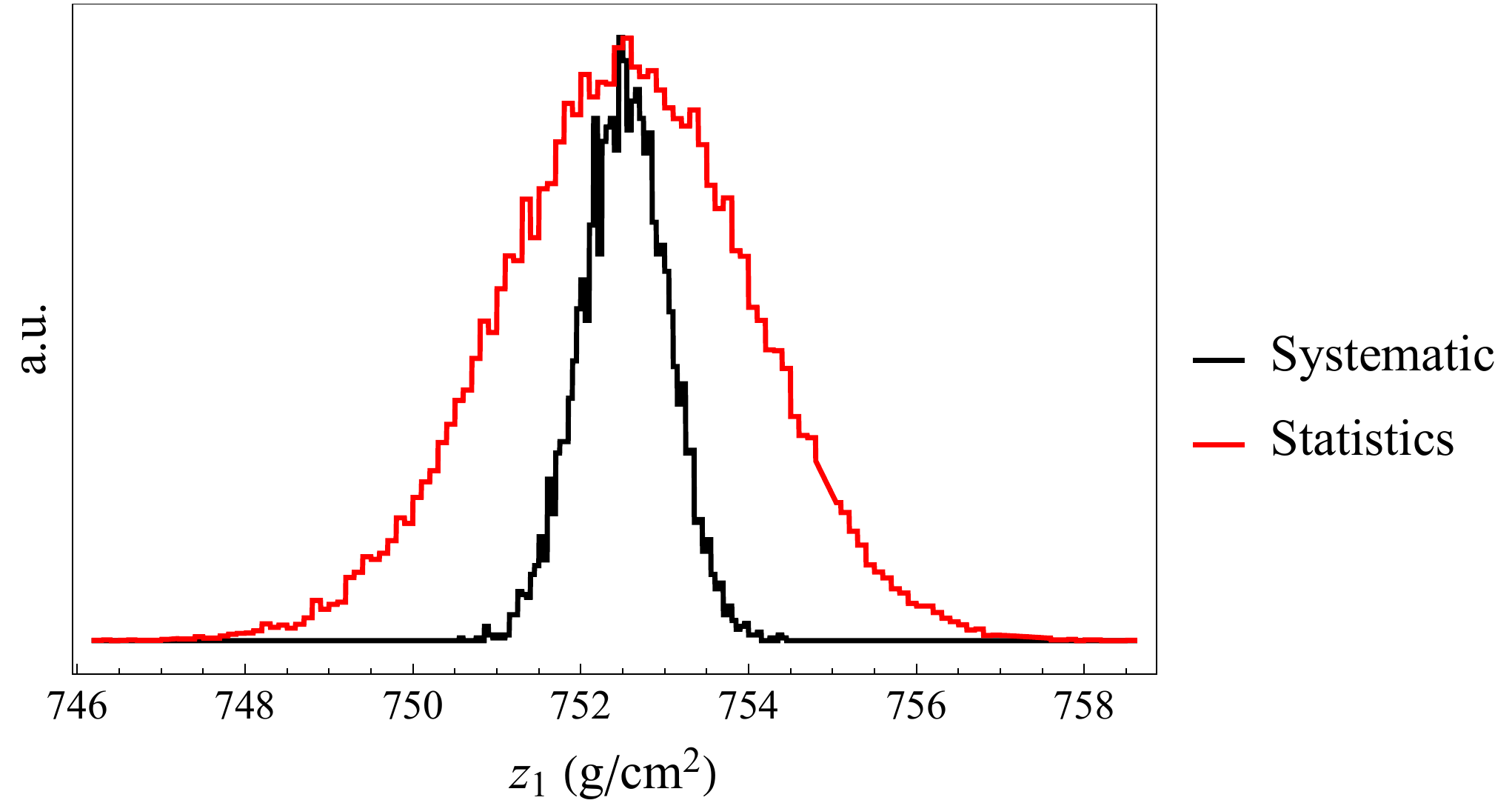}
        \caption{Distribution of $z_1$ for protons simulated with EPOS in the $[1,2]$ EeV energy bin. The black distribution is obtained including only systematic uncertainties, while the red one including statistical uncertainties by bootstrapping, see Sec.~\ref{subsec:Bootstrapping} for details. 
        }
        \label{fig:Syst_vs_Stat}
\end{figure}

The systematic uncertainty of a single shower $\Xm^j$ is fully described by the modified distribution $F_j(X|Z)$, see Eq.~\eqref{eq:Fnotation}. The mean and standard deviation of the latter, namely $\mu_j = {\cal F}_j^1(Z)/{\cal F}_j^0(Z)$ and $\sigma_j = \sqrt{{\cal F}_j^2(Z) - [{\cal F}_j^1(Z)]^2}/{\cal F}_j^0(Z)$, can be seen as the best estimations of the $j$-th simulated measurement and its uncertainty. To show the total contribution of the systematic error, we sample each simulated shower from a normal distribution ${\cal N}\lp \mu_j,\sigma_j\rp$, and compute the central moments of the resulting distribution. By repeating these steps multiple times, we obtain an estimation of each moment distribution, and in particular their width. We show this in Fig.~\ref{fig:Syst_vs_Stat} (in black) for the example case of the first moment of proton showers simulated with EPOS, in the energy bin $[1,2]$ EeV. 

The statistical uncertainties of moments $z_n$ are estimated in a similar way using bootstrapping.
We sample an event $N$ times from a set of $N$ showers, allowing for repetitions and with the same probability $1/N$ to be picked. Sampled events are then used to compute moments $z_n$. By repeating this procedure multiple times we obtain a set of moments $\{z_n\}$ which captures statistical and systematic uncertainties at the same time.
To have a more intuitive picture, we can think of the infinite statistics limit, $N\to\infty$, while neglecting for this purpose any systematic error. In this limit, the probability of sampling any single event more than once goes to 0; the resulting distribution of each moment will tend to a delta function peaking at the real value of $z_n$. Reducing the number of events available, we are more likely to sample the same event multiple times, resulting in a wider distribution of $z_n$.

We indicate the steps of the bootstrapping procedure with the index $l$. The PDF of the $l$-th bootstrapped sample is
\beq
\label{eq:bootstrappedPDF}
P_{{\rm sim}} (X ~|~ Z)_l = \frac1N \sum_j{\cal O}_{j,l}(Z) F_j (X ~|~ Z)\,,
\eeq
where ${\cal O}_{j,l}(Z)$ gives the number of times the $j$-th event is sampled in the $l$-th bootstrapped step. At each $l$ we generate a random list ${\cal O}_{j,l}(Z)$, with the single constraint that $\sum_j {\cal O}_{j,l} = N$. 
The $n$-th moment for a given composition $w$, Eq.~\eqref{eq:Wmoments}, at the $l$-th bootstrapped step then reads
\beq \label{Eq.:noncentral_main}
\langle \Xm^n \rangle_l(w) = 
\frac{\sum_Z G(Z)^n_l ~w_Z}{\sum_Z G(Z)^0_l ~w_Z},
\eeq
where we have defined
\beq
G(Z)^n_l \equiv \frac{1}{N}\sum_j{\cal O}_{j,l}(Z){\cal F}_j^n(Z)\,.
\eeq
This expression summarises our notation of Sec.~\ref{subsec:MomentDecomposition}, as for $n = 0$ we have $G(Z)^0_l = \Delta_{Z,l}$ and for
$n > 0$ we get $G(Z)^n_l = \langle X^n_{\text{max}} \rangle_{Z,l} ~\Delta_{Z,l}$. All information about simulated showers used to infer the composition is then entailed by the tensor $G(Z)^n_l$.

We perform the bootstrapping procedure with $M = 10^5$ steps. For each step we compute the moments $z_n$ using Eq.~\eqref{eq:CentralMoments}. This results in a set of moments, $\{ (z_n)_1,\dots,(z_n)_M \}$, which we use to obtain their distributions.
As a case example, we show in Fig.~\ref{fig:Syst_vs_Stat} (in red) the result for the $z_1$ distribution of proton showers, simulated with hadronic model EPOS in the $[1,2]$ EeV energy bin.
It is clear from Fig.~\ref{fig:Syst_vs_Stat} that the total uncertainty of the central moment $z_1$ is dominated by statistical fluctuations due to finite number simulated showers, which dominate over the systematic errors from detector effects. In the example shown, the width of the $z_1$ distribution is $\sim 0.5~\gcm$ for the latter, while it is  $\sim 1.5~\gcm$ for the former. That is, including statistical errors increases the total uncertainty by a factor of 3. The same pattern can be seen for the other moments and in general for all other primaries, where the ratio between total and systematic errors is closer to a factor of 2 for our simulated shower samples. 

We can apply the same procedure to the set of measured data. Starting from the PDF for measured $\Xm$ in a fixed energy bin, Eq.~\eqref{eq:PDF:Auger}, we can write at the $l$-th step of the bootstrapping
\beq \label{eq:PDF:Auger:Bootstrapped}
P(X~|~E)_l = \frac1N\sum^N_{j = 1} \mathcal{N}(X ~| ~X_{\max,j}, ~\delta X_{\max,j}) ~\mathcal{O}_{j,l},
\eeq
where now $N$ is the number of measured events in bin $E$. It follows that the $n$-th moment is
\beq\label{eq:Moments:Auger:Bootstrapped}
\langle \Xm^n \rangle_l = \int P(X~|~E)_l ~X^n ~{\rm d}X 
= 
 \frac1N\sum^N_{j = 1} \mathcal{O}_{j,l} \int \mathcal{N}(X ~| ~X_{\max,j}, \delta X_{\max,j}) X^n ~{\rm d}X\,.
\eeq
Similarly to the case of Eq.~\eqref{eq:calF}, we can compute these integrals for each event once, before performing the bootstrapping, thus greatly improving on the required computation time. 

\begin{center}
\begin{figure}[!t]
        ~~~~~\includegraphics[width=0.9\linewidth]{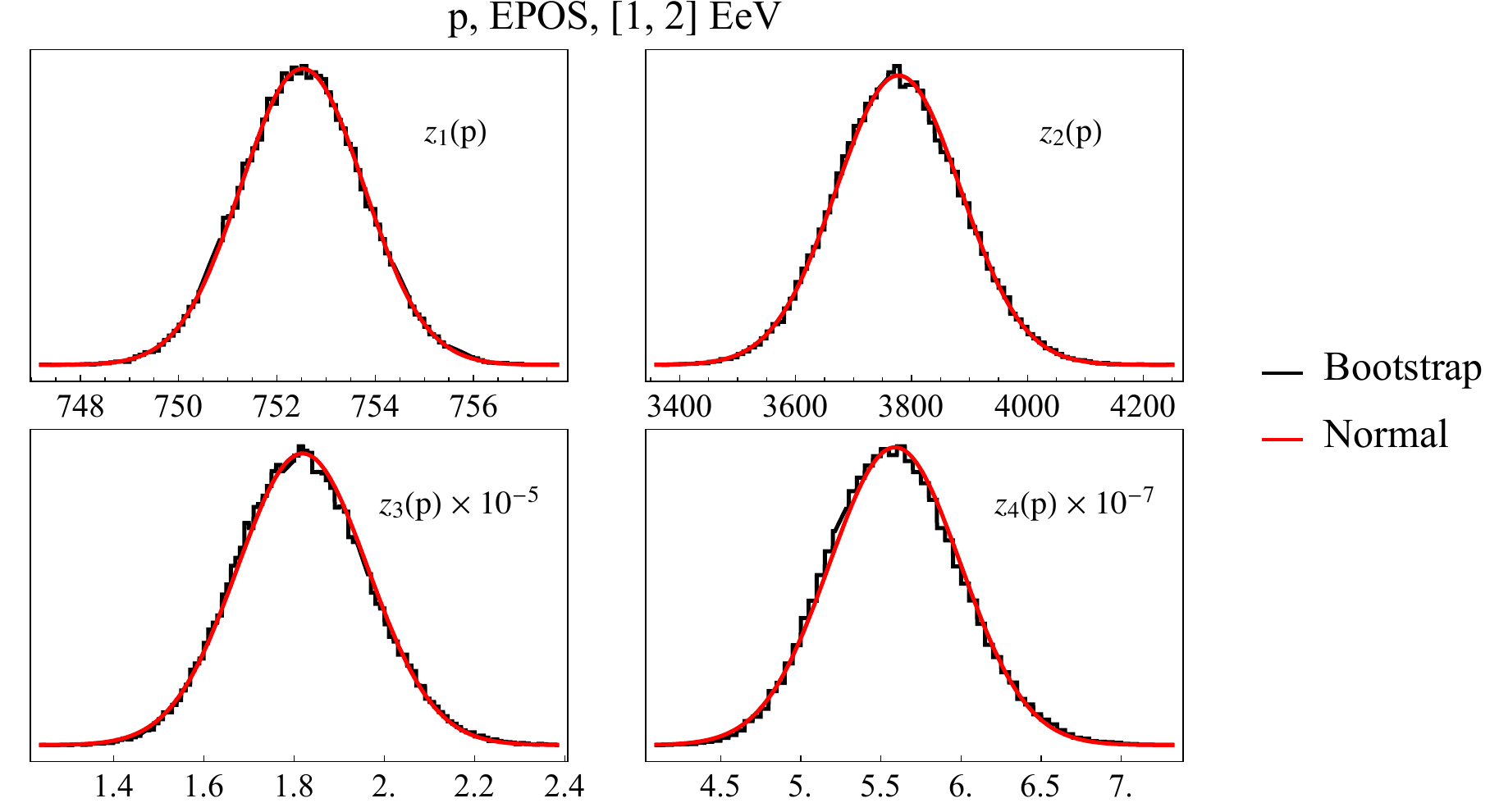}
        \caption{Distribution of first four moments, $z_{1,2,3,4}$, for protons simulated with EPOS in the $[1,2]$ EeV energy bin. The horizontal axis of the $z_n$ plot is in units $(\gcm)^n$. The black line indicates the distribution of the respective moment obtained by bootstrapping the simulated $\Xm$. The mean and standard deviation of the latter define the normal distribution shown with the red line.}
        \label{fig:Normal_vs_BS}
\end{figure}
\end{center}

Finally, we remark that the bootstrapped moments closely follow normal distributions, as expected from the central limit theorem, both for simulations and real data. In Fig.~\ref{fig:Normal_vs_BS} we show the first four moments distributions for protons simulated with EPOS in the $[1,2]$~EeV energy bin. 
In the following we can then safely take $p(z_i) = {\cal N}(z_i~|~\mu_{z_i},\sigma_{z_i})$, where $(\mu_{z_i},\sigma_{z_i})$ are the mean and standard deviation of $\{ z_i \}$. More generally we have 
\beq\label{eq:Moments:Multivariate}
z(w) \sim {\cal N}_n\Big( z ~|~ \mu(w), ~\Sigma(w)\Big)\,,
\eeq
where $z(w)$ is the vector of $n$ moments and ${\cal N}_n(z ~|~ \mu, \Sigma)$ is a multivariate normal distribution, with $\mu$ the $n-$dimensional mean vector and $\Sigma$ the $n\times n$ covariance matrix. As described in Eq.~\eqref{eq:Wmoments}, in the general case the moments of $\Xm$ distributions will depend on the composition; here this is reflected in $\mu$ and $\Sigma$ being functions of $w$.

Similarly, from Eq.~\eqref{eq:Moments:Auger:Bootstrapped} we have for Auger data in a selected energy bin
\beq\label{eq:Moments:Multivariate:Auger}
\tilde z \sim {\cal N}_n\lp z ~|~\tilde\mu, \tilde\Sigma \rp\,,
\eeq
where now the mean $\tilde\mu$ and covariance matrix $\tilde\Sigma$ are constants calculated from the (bootstrapped) data distributions for each energy bin.

\subsection{Likelihood}
\label{subsec:Composition:Likelihood}

The composition $w$ of UHECR in a selected energy bin is inferred by comparing simulated data with the GH events in the Open Data from Pierre Auger Observatory. In the following discussion, we assume that both the energy bin and the hadronic model have been fixed, thus leaving the composition $w$ as the only free parameter.

In the previous sections we mapped the PDFs of both measured and simulated $\Xm$ to a set of $n$ features, namely the first $n$ central moments of the distributions, via Eq.~\eqref{eq:DimRedMap}. Furthermore, we have shown that we can safely approximate the moment distribution in terms of a $n$-dimensional multivariate normal. For a given composition $w$, the moments of simulated data can be expressed as a weighted average of single primary moments, as described in Eq.~\eqref{eq:Wmoments}. Thus the parameters of the respective multivariate distribution in Eq.~\eqref{eq:Moments:Multivariate}, the mean vector $\mu(w)$ and the covariance matrix $\Sigma(w)$, contain all the information on the composition $w$. Similarly, the distribution of moments of measured data is described by the parameters $\tilde\mu$ and $\tilde\Sigma$, as in Eq.~\eqref{eq:Moments:Multivariate:Auger}.

Given the above premises, the problem of inferring the composition consists of fitting the simulated $n$-dimensional vector $z(w)$, to the measured $\tilde{z}$. The approximation to multivariate normal distribution further simplifies the problem of taking into account the full uncertainties and correlations into the likelihood function. In particular, the likelihood of obtaining moments $z$, with composition $w$, given the experimental data  reads
\beq
\label{eq:tildeL}
    \tilde{\cal L}(z, w) = {\cal N}_n\Big( z ~|~ \tilde\mu, \tilde\Sigma \Big) \times {\cal N}_n\Big( z ~|~ \mu_w, \Sigma_w\Big)\,,
\eeq
where we have written $\mu(w)\equiv\mu_w$, $\Sigma(w) \equiv \Sigma_w$, for brevity. In general we are considering the likelihood as a function of both $z$ and $w$. The former can be treated as a nuisance parameter, introduced to take into account the uncertainties, which we need to marginalize over. 
In this work we follow the Bayesian approach and integrate Eq.~\eqref{eq:tildeL} over all possible values of $z$
\beq
\label{eq:LikelihoodGeneral}
{\cal L}(w) = \int\tilde{\cal L}(z, w) ~{\rm d}^n z\,.
\eeq
The integral can be solved explicitly, and the logarithm of the solution (the log-likelihood) reads
\beq\label{eq:LogLikelihood}
\begin{split}
\log\left[{\cal L}(w)\right] =& -\frac{n}{2} \log(2 \pi) - \frac12 \log\left[ \det(\Sigma_w + \tilde\Sigma)\right] - \frac12 \lp \mu_w^T \Sigma_w  \mu_w +   
\tilde\mu^T \tilde\Sigma \tilde\mu \rp \\
&+ \frac12
\lp \Sigma_w^{-1}\mu_w + \tilde\Sigma^{-1}\tilde\mu  \rp^T
\lp \Sigma_w^{-1} +  \tilde\Sigma^{-1} \rp^{-1}
\lp \Sigma_w^{-1}\mu_w + \tilde\Sigma^{-1}\tilde\mu  \rp\,.
\end{split}
\eeq
Finally, we can obtain the distribution of possible compositions $w$,  ${\cal P}(w)$, given the experimental data, via the Bayes theorem as
\beq\label{eq:Posterior}
    {\cal P}(w) = \frac{{\cal L}(w) ~{\rm Dir}(w, \alpha)}{ \int {\cal L}(w) ~{\rm Dir}(w, \alpha)~ {\rm d}^D w}\,,
\eeq
where $D=26$. Here we have assumed a flat prior (Dirichlet) distribution ${\rm Dir}(w, \alpha)$, with $\alpha = (1,...,1)$, meaning that in the absence of experimental information, all compositions are equally probable. The ``best composition'', $w^*$, is thus the composition that maximizes the posterior ${\cal P}(w)$ or equivalently in case of a flat prior, the composition that maximizes the log-likelihood in Eq.~(\ref{eq:LogLikelihood}). 

The posterior distribution in Eq.~\eqref{eq:Posterior} realizes the main goal of this work: the most probable composition as well as the confidence regions (or confidence intervals of single primary fractions), can be extracted from an unbinned likelihood of $\{\Xm\}$ distributions based on their expansion in central moments.

\subsection{Estimating Confidence Regions with Nested Sampling}
\label{subsec:Composition:Method}

While the best composition $w^*$ can be obtained in a straightforward way, by maximizing the  log-likelihood in Eq.~\eqref{eq:LogLikelihood}, the extraction of confidence intervals is significantly more involved. The likelihood ${\cal L}(w)$ depends on 26 correlated parameters, subject to a single constraint $\sum_Z w_Z = 1$. The numerical evaluation of such a function around the point $w^*$ proves to be computationally intensive. We approach this problem by sampling from the posterior ${\cal P}(w)$ using a Nested Sampling (NS)~\cite{doi:10.1063/1.1835238} algorithm. 

The basic task of NS is to compute the evidence
\begin{equation}
    Z = \int {\cal L}(w) ~{\rm Dir}(w) ~{\rm d}^Dw = \int^1_0 {\cal L}(X) ~ {\rm d}X\,,
\end{equation}
where ${\cal L}(X)$ is obtained by inverting the mass function $X(L) = \int_{{\cal L}(w) \geq { L}} ~\text{Dir}(w) ~{\rm d}^Dw$. 
In its simplest form, the algorithm can be described in the following way. In the first step, $k = 1$,
$N_{\text{live}}$ compositions $w$, called live points, are sampled from the prior. The point $w_1$ with the lowest likelihood, $L_1$, is called a dead point. At each following step $k > 1$ a new live point is sampled from the prior with the constraint that $\mathcal L(w) > L_{k-1}$ and again a dead point $w_k$ with likelihood $L_k$ is determined.
The contribution to the evidence at step $k$ is given by $\delta Z_k = L_k ~\delta X_k$, where $\delta X_k$ is the the volume of the prior region where points have likelihood between ${L}_{k-1} < \mathcal{L}(w) \leq {L}_k$; this region can be estimated from a Beta distribution for uniform priors (see Ref.~\cite{doi:10.1063/1.1835238} for details). Finally, the algorithm outputs a set of dead points, $w_k$, with the associated weights, $u_k$, given by $u_k = \delta Z_k / Z$, where $k = 1,\dots,M $ are the number of samples produced. The size of $M$ depends on the number of $N_{{\rm live}}$ employed in the sampling procedure. 

Built on this simple procedure, the modern implementations of the NS algorithm include in addition evaluations of the uncertainty on the estimated volume of the prior region as well as the uncertainty on the contribution to the evidence $\delta Z_k$. The procedure is in principle valid for any prior, which in our case is assumed to be a flat Dirichlet distribution.

\begin{figure}[t!]
    \centering
    \includegraphics[width=.8\linewidth]{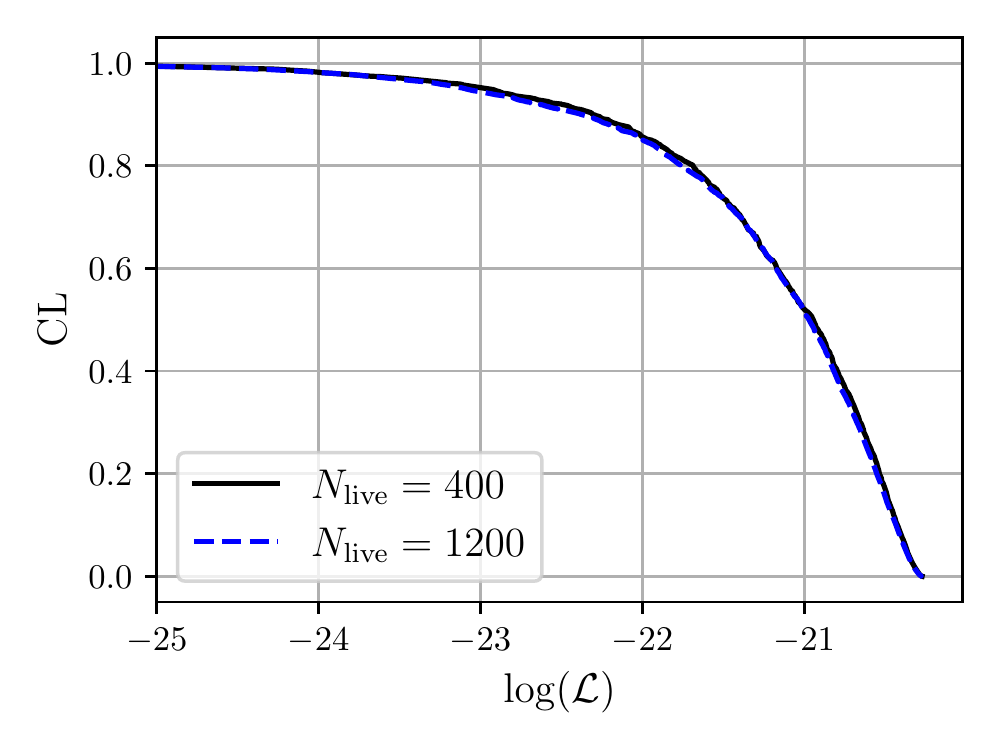}
    \caption{Confidence level as a function of Log-Likelihood, for hadronic model EPOS in the energy interval $17.9 < \log(E/\text{eV}) \leq 18.0$. The solid black line shows the result with $N_{{\rm live}} = 400$, while the dashed blue line is for $N_{{\rm live}} = 1200$. In both cases we use $n=3$ moments as features. 
    }
    \label{fig:CLlogL}
\end{figure}

In this work we use a recent implementation of NS, called {\tt UltraNest} \cite{Buchner_2014, https://doi.org/10.48550/arxiv.1707.04476, https://doi.org/10.48550/arxiv.2101.09604}, available on Github~\cite{UNgithub}. The computation of weighted samples from the posterior is done using the function {\tt ReactiveNestedSampler}. We also employ the slice sampling technique, included in the {\tt UltraNest} code, with the default setting for the number of steps. The latter allows to efficiently explore the high dimensional space spanned by our parameter.
We then use the outputs $\{(w_k, u_k) \}$ to compute the confidence level as a function of likelihood ${\cal L}_0$ as
\begin{equation}
\label{eq:CL} 
    \text{CL}({\cal L}_0) = \sum_{(w_k,u_k) ~|~ {\cal L}(w) \geq {\cal L}_0} u_k\,.
\end{equation}
In Fig.~\ref{fig:CLlogL} we show the latter relation obtained for the 4 primaries mixture described in Sec.~\ref{subsec:Comparison:4primaries}, with two different settings for the number of live points used by {\tt UltraNest}, $N_{{\rm live}} = 400$ and $N_{{\rm live}} = 1200$. As the two results are consistent, we use the lower $N_{\rm live}=400$ setting, which considerably reduces the computation time required. 

\begin{figure}[h!]
    \centering
    \includegraphics[width=.7\linewidth]{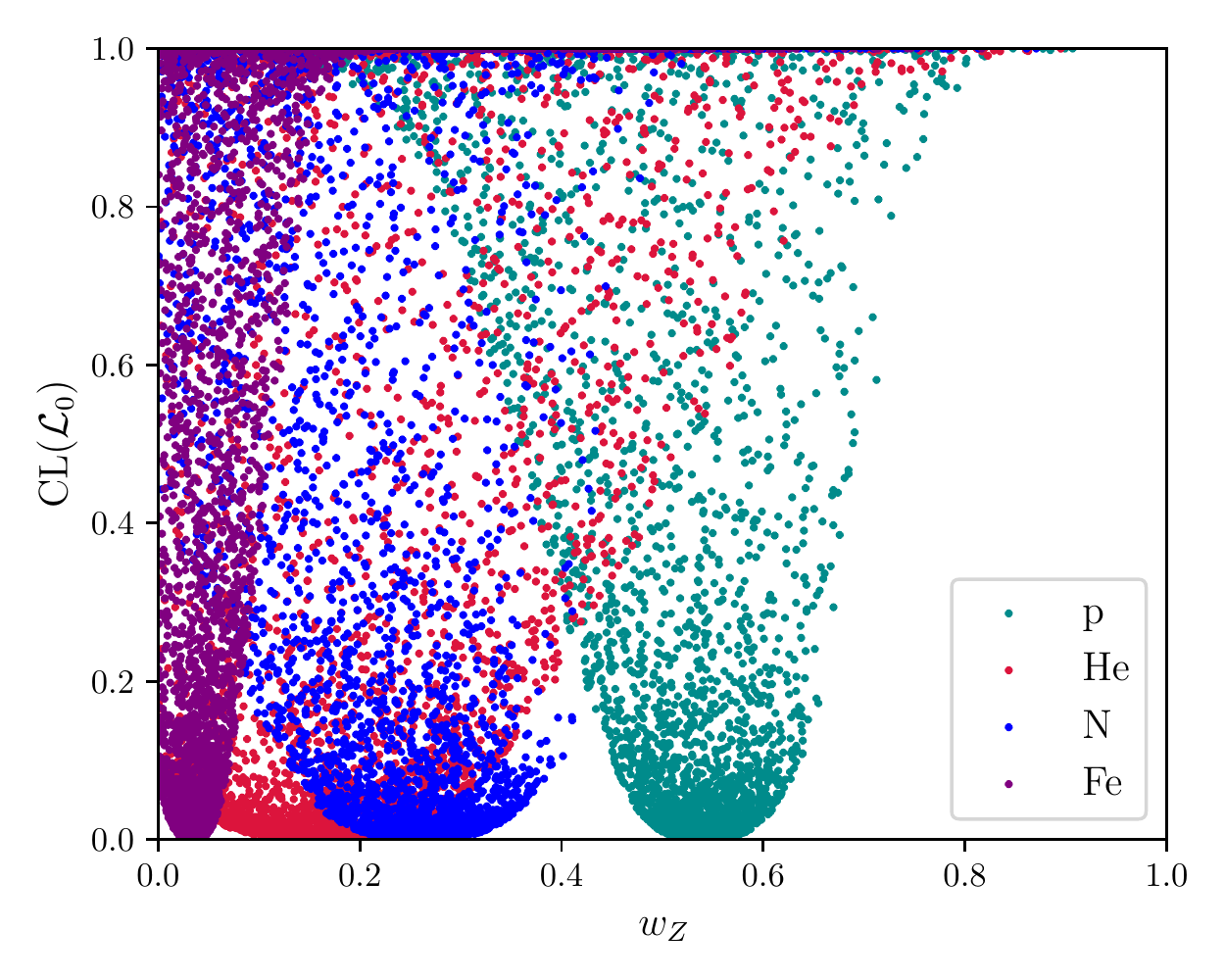}
    \caption{Fraction of primaries from samples generated by NS with 400 live points, for hadronic model EPOS in the energy interval $17.9 < \log(E/\text{eV}) <18.0$. Here we use $n=3$ moments.}
    \label{fig:scatter_wZ}
\end{figure}

Samples generated by NS cannot be used directly to determine the confidence regions of individual primary fractions, $w_Z$, as can be seen from Fig.~\ref{fig:scatter_wZ}. In general the algorithm does not provide samples of $w$ that lie precisely on the boundary of a confidence region of interest. Instead, it provides a reliable map between the confidence level $\text{CL}(\mathcal{L})$
and likelihood $\mathcal{L}$, as shown in Fig.~\ref{fig:CLlogL}. With this information we can compute the confidence intervals for a fixed $\text{CL}({\cal L}_0)$ and primary $Z$ by solving for the positivity limits of the function ${\cal L}(w_0) - {\cal L}_0$. That is, we look for the two compositions $w_0^{{\rm low}}$ and $w_0^{{\rm high}}$, which satisfy the following condition
\beq 
\forall~ w_0 ~: ~{\cal L}(w_0) \geq {\cal L}_0 ~~ \Rightarrow ~ 
(w_0)_Z~\in~[(w_0^{{\rm low}})_Z,~(w_0^{{\rm high}})_Z]\,,
\eeq
 where {$(w_0)_Z$ is the $Z$-th component}.
Specifically, we compute the upper and lower bounds for each primary fraction for the $68.3\%$ ($1\sigma$) and $95.4\%$ ($2\sigma$) confidence levels separately. Starting with the lower bound, we describe here the algorithm for a fixed primary $Z$ and confidence level ${\rm CL}({\cal L})$:
\begin{enumerate}
    \item Choose the confidence level $\rm CL$ (e.g., $1\sigma$) and calculate the corresponding $\mathcal{L}_0$ by inverting the relation in Eq.~\ref{eq:CL}.
    \item Select 3 compositions $w_0$ from samples obtained by NS algorithm, which have the lowest values of $Z$-th component $(w_0)_Z$ and with likelihoods $\mathcal{L}(w_0) \geq {\cal L}_0$.
    \item For each selected composition $w_0$ initialize $n = 0$ and repeat the following steps:
    
    \begin{itemize}
        \item[a)] Sample $M = 200$ times from a Multinomial distribution $\text{Mult}(w| ~p = w_0, ~N = 1000)$. 
        In this way a set of compositions $\{ w_1,..., w_M\}$ is obtained.
        \item[b)] Find a composition $w^*$ in the set $\{ w_1,..., w_M\}$ with likelihood ${\mathcal{L}(w^*)} \geq {\cal L}_0$ and with the lowest value of the Z-th component, $(w^*)_Z$.
        \item[c)] If $(w^*)_Z < (w_0)_Z$ replace $w_0$ with $w^*$ and set $n = 0$, otherwise add 1 to $n$.
        \item[d)] If $n = 10$ exit the loop. The value of $(w^*)_Z$ is the estimated lower bound for the confidence interval of primary $Z$ for confidence level $\rm CL$. Save this value.
    \end{itemize}

    \item Compare estimated lower bounds and determine the lowest value among the three.\footnote{The estimated bounds do not improve significantly by using a higher number of compositions in step 2. We use 3 points to reduce the computational time and check the consistency of the results.}
\end{enumerate}

The upper bound and the most probable composition can be found using the same approach with trivial modifications. In Fig.~\ref{fig:scatter_wZ} we show the result for the single primary fractions, obtained in the 4 element mixture case described in Sec.~\ref{subsec:Comparison:4primaries}, for different confidence levels.  

%
\section{Results}
\label{sec:Composition:Results}
%

\subsection{Method Validation and Comparison with Previous Studies}
\label{subsec:Comparison:4primaries}

We first apply our method to a mixture of 4 elements, namely (p,~He,~N,~Fe), in order compare with results of previous studies~\cite{PierreAuger:2014gko, Arsene:2020ago}. 
In particular, in these works, the composition of up to 8 primaries (p, He, C, N, O, Ne, Si, Fe) was inferred by binning the $\Xm$ distribution, both simulated and measured, and maximizing the likelihood
\beq
\label{eq:ProfileLikelihood}
    \log\left[\mathcal{L}_{\rm bin}(w)\right] = \sum^N_{i = 1} \left( n_i - y_i  - n_i \log\left[n_i/y_i\right]\right)\,, 
\eeq
where $n_i$ is the number of simulated showers in the $i$-th bin of $\Xm$ and $y_i$ is the number of observed events in the same bin, $N$ is the number of bins. 

\begin{figure}[!t]
    \centering
    \includegraphics[width=.45\linewidth]{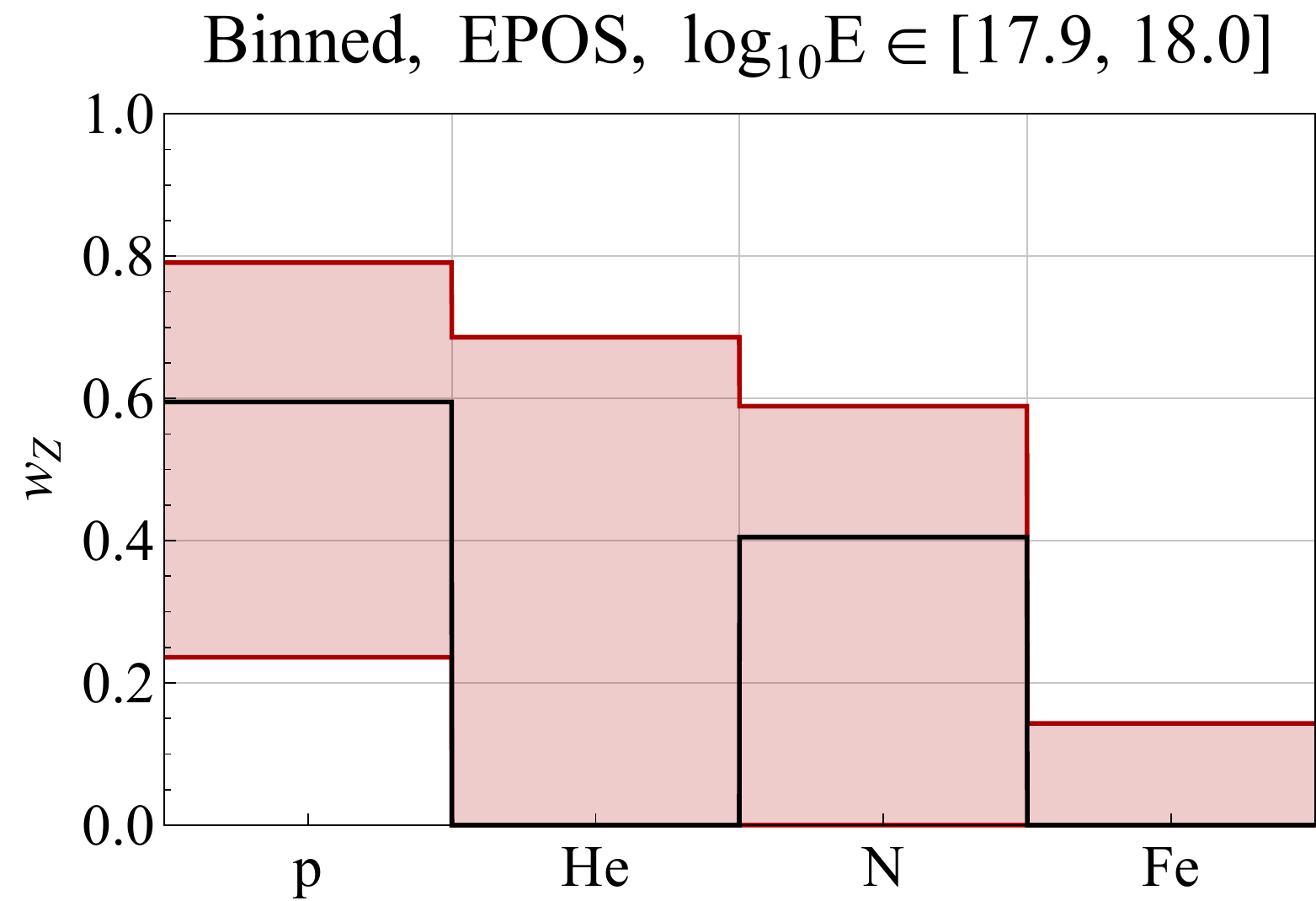}\hspace{1cm}
    \includegraphics[width=.45\linewidth]{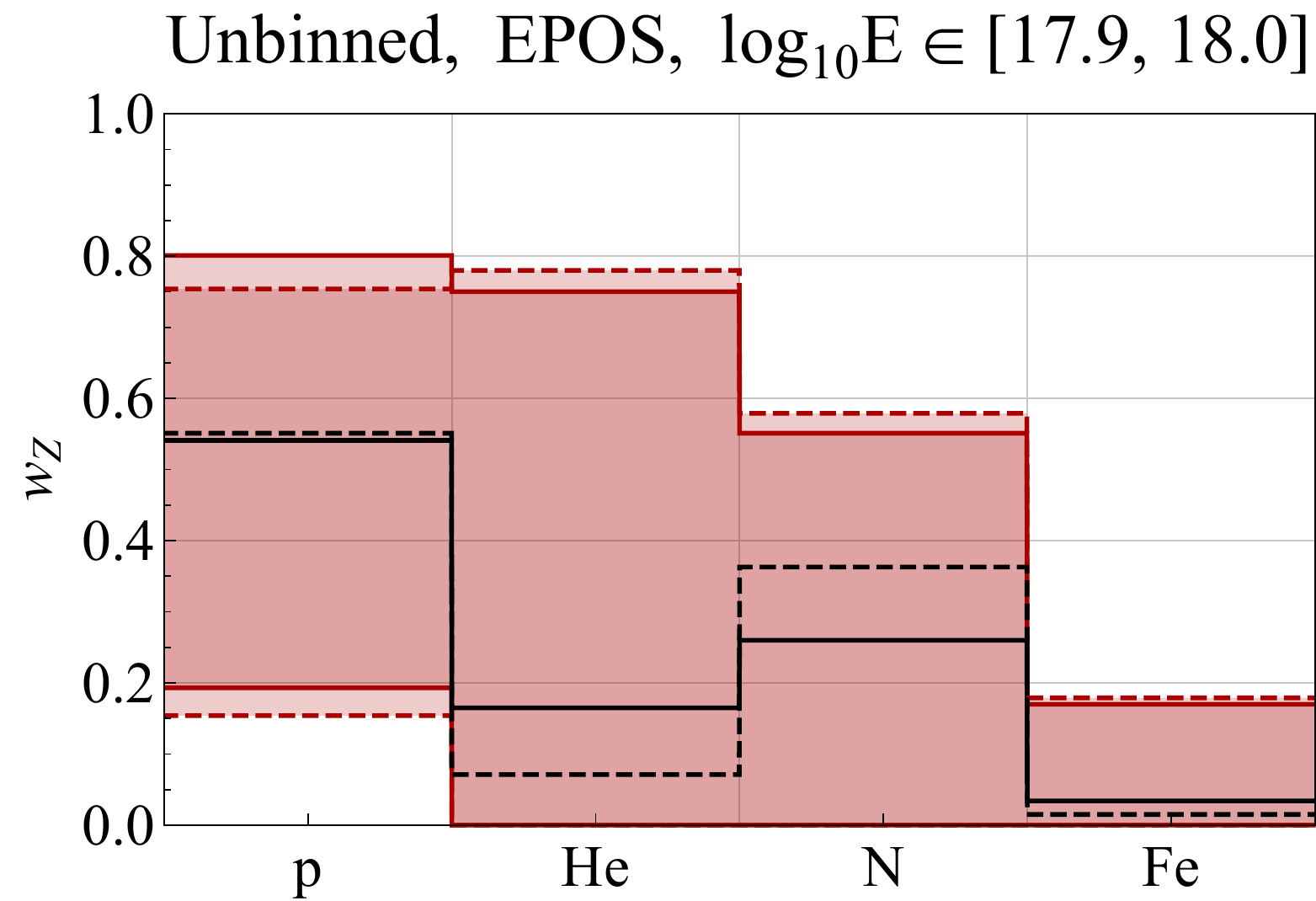}
    \caption{Inferred composition for a mixture of 4 primaries, (p, He, N, Fe), in the energy bin $\log_{10}E\in[17.9,~18.0]$. {\bf Left:} result obtained using the binned likelihood, Eq.~\eqref{eq:ProfileLikelihood}. The black solid line indicates the most probable composition, while the red bands the $2\sigma$ confidence interval. {\bf Right:} results obtained with the unbinned likelihood, Eq.~\eqref{eq:LikelihoodGeneral}. The black and red solid lines show the best composition and the $2\sigma$ regions respectively, based on $n=3$ central moments. The dashed lines represent the same in the case of $n=4$ central moments.
    }
    \label{fig:4primaries:Results1and2sigma}
\end{figure}

In Fig.~\ref{fig:4primaries:Results1and2sigma} we compare the results, in the framework of 4 primaries in the $\log_{10}(E/ \rm eV)\in[17.9,~18.0]$ energy bin, obtained with the use of the binned likelihood with $N=46$, Eq.~\eqref{eq:ProfileLikelihood} (left plot), and with the use of the unbinned likelihood for the $\Xm$ central moment decomposition with $n=3,4$, Eq.~\eqref{eq:LikelihoodGeneral} (right plot). 
Both methods have been applied to the same available events from the Auger Open Data set, resulting in somewhat wider confidence intervals compared to the original publications~\cite{PierreAuger:2014gko, Arsene:2020ago}. Nonetheless, the general features of the results in the literature are easily reproduced in our analysis, with a preference for large proton fractions, $w_p\sim60\%$ for the best fit, but with a sizeable presence of heavier elements, i.e. $w_N\sim40\%$. Importantly, the inferred fraction confidence intervals are comparable between the two approaches, even though the unbinned method is based only on the first three or four central moments of the $\Xm$ distribution compared to $N=46$ bins considered in the binned likelihood approach. Note also that results of unbinned fit with $n=3,4$ are well consistent with each other. Small observed shifts in the $w^*$ and its confidence intervals can be expected given the low statistics and high dimensionality of the problem. Importantly however the general features are not changed. 

\begin{figure}[!t]
    \centering
\includegraphics[width=0.8\linewidth]{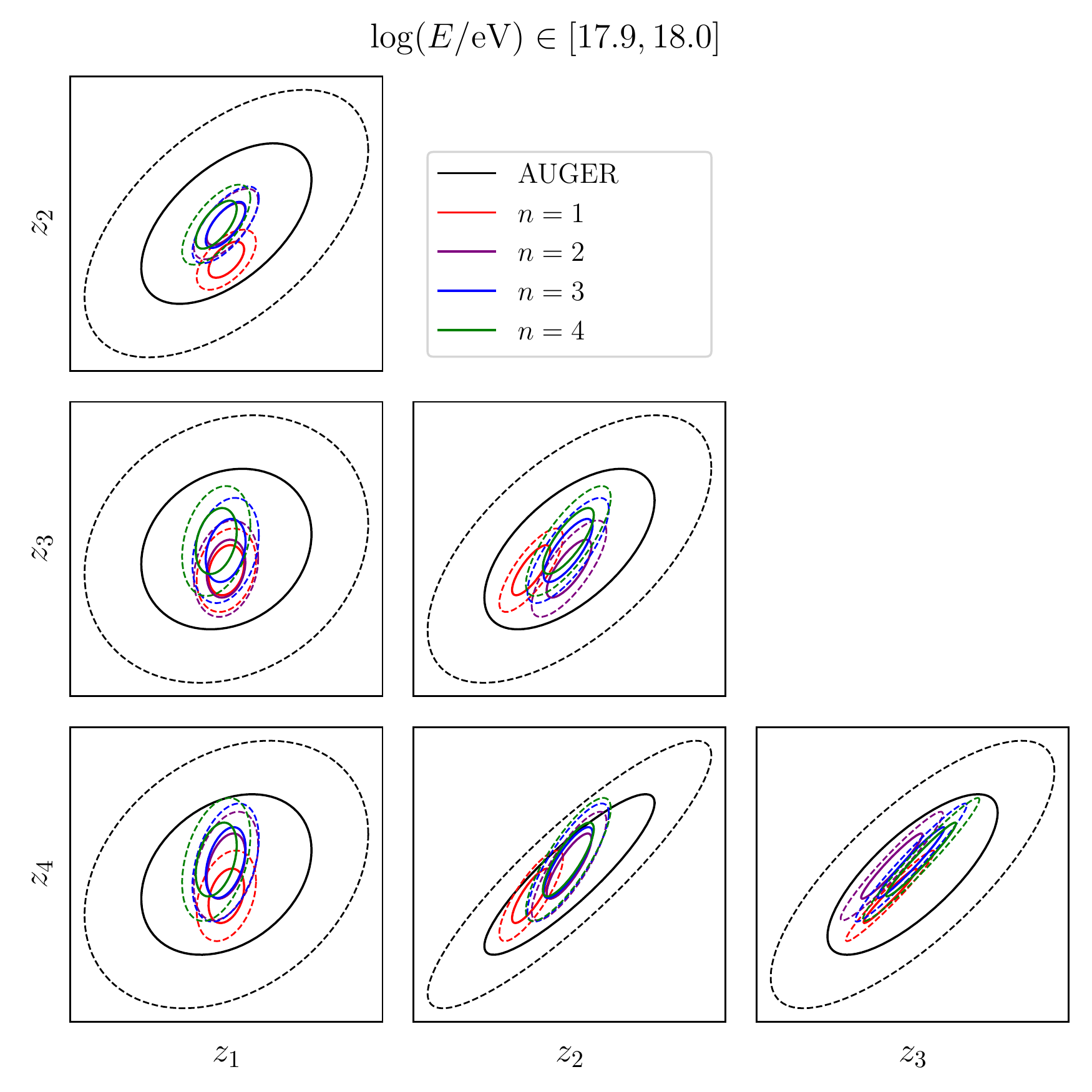}
\caption{$3\sigma$ (solid) and $5\sigma$ (dashed) contours of the multivariate moment distributions for the Auger data (black) and for the best composition inferred with EPOS, using $n=1$ (red), $n=2$ (purple), $n=3$ (blue) and $n=4$ (blue) moments, in the $\log_{10}E\in[17.9,~18.0]$ energy bin, for the 4 primary framework.}
\label{fig:4Composition:3vs4moments:LowBin:z1-z4:EPOS}
\end{figure}

We further explore the qualitative and quantitative differences between results obtained with fits to different numbers of central moments in Fig.~\ref{fig:4Composition:3vs4moments:LowBin:z1-z4:EPOS}. 
We show the $3\sigma$ (solid) and $5\sigma$ (dashed) contours of the multivariate normal distributions of moments in the same energy bin considered above, $\log_{10}E\in[17.9,~18.0]$, projected onto planes spanned by pairs of moments, see eq.~\eqref{eq:Moments:Multivariate} and \eqref{eq:Moments:Multivariate:Auger}. The black ellipses indicate the contours of the measured Auger Open Data moment distribution, while the red, purple, blue and green lines indicate the maximum likelihood compositions, inferred from unbinned fits to $n=1,2,3,4$ central moments, respectively. Note that when fitting to only $z_1 = \langle \Xm \rangle$, the best fit composition gives rather poor predictions for higher moments. This again indicates that higher central moments of the $\Xm$ distribution contain additional relevant additional information on the primary composition of UHECRs. On the other hand, when including the second (and third) moments in the fit, the resulting predictions for the higher (third (and fourth)) moments are consistent within uncertainties with the data, but more importantly, also with model results obtained when these higher moments are included in the fit. This reaffirms our expectation that (given the available statistics) $n=3$ is sufficient to describe the most relevant features of the $\Xm$ distributions when inferring the composition of UHECRs.\footnote{We revisit this issue again in Sec.~\ref{subsec:Fullcomposition:Results} when considering the full $Z=26$ composition results.}

\subsection{Full Composition Results - EPOS}
\label{subsec:Fullcomposition:Results}

Next we apply our method to infer the full composition of UHECRs. In this subsection we focus on results obtained with the EPOS hadronic model. We compare results obtained with different hadronic models in Sec.~\ref{subsec:HMComaprison:Results}. Additional plots and results are collected in Appendix~\ref{app:extraplots}.

\begin{figure}[!t]
    \centering
\includegraphics[width=.7\linewidth]{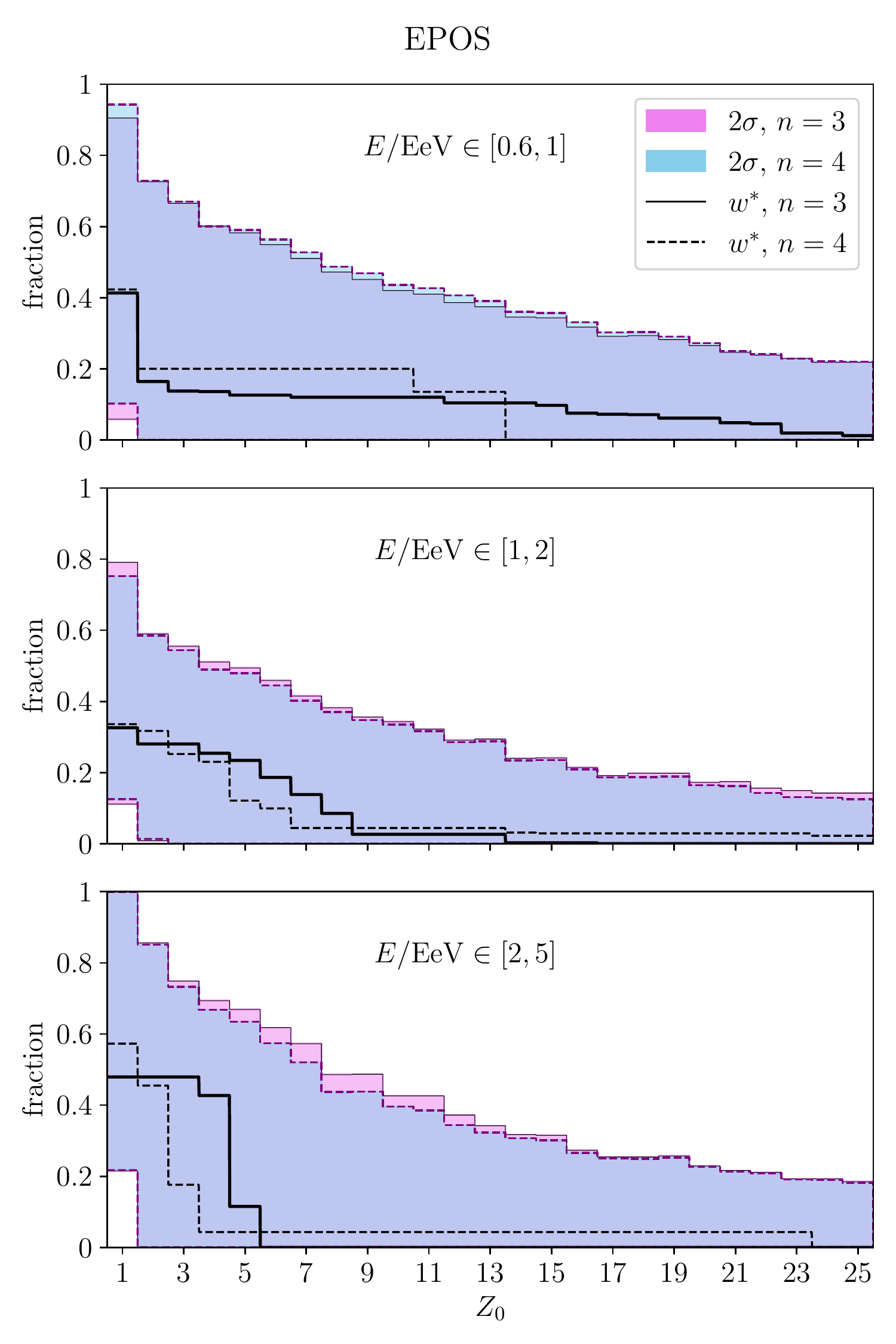}
\caption{Fraction of primaries with atomic number $Z > Z_0$ inferred with the EPOS model, in the three energy bins considered. The solid black line and purple regions indicate the results taking $n=3$ moments in the distribution decomposition, while the dashed black line and cyan regions show the effect of taking $n=4$ moments.}\label{fig:FullComposition:3vs4moments:Cumulative:EPOS}
\end{figure}

To show the inferred compositions of all 26 considered primaries in a meaningful way, while properly including the confidence intervals and account for correlations, we plot cumulative fractions of elements. That is, for each $Z_0\in\{1,...,~26\}$, we plot the fractions of elements heavier than $Z_0$ ($Z>Z_0$) that form the showers, with the respective $1\sigma$ and $2\sigma$ confidence intervals. The single fractions $w_Z$ suffer from the fact that the full composition $w$ represents a point in a $26$ dimensional space, with a single constraint $\sum_Z w_Z = 1$. Thus, the confidence interval for any single $w_Z$, which is  actually a projected confidence region of $w$ on the $Z$-dimension, does not carry useful information on the remaining 25 fractions. Furthermore, the low available statistics leads to very large confidence intervals of the inferred fractions. The value of any single $w_z$, while interesting on its own, cannot be strongly constrained with presently available Open Data.

In Fig.~\ref{fig:FullComposition:3vs4moments:Cumulative:EPOS} we show the results obtained in the three energy bins with EPOS. The black (dashed) lines indicate the cumulative for the best composition $w^*$, obtained by maximizing the posterior probability, Eq.~\eqref{eq:Posterior} with $n=3 (4)$, while the magenta (blue) shaded regions indicate the respective $2\sigma$ confidence intervals. At each step $Z_0$ we can read the fraction of elements heavier than $Z_0$. 

Focusing on the confidence of exclusion indicated by the $2\sigma$ regions, at the level of precision achievable with the Open Data, we can exclude that $\gtrsim90\%$ of the showers are sourced by protons. {Or, in other words, that at least $\sim10\%$ of the showers are sourced by elements heavier than the  proton.} Similar limits can be extracted for all elements and energy bins. Despite the low precision of such predictions, it can still be seen how higher energy showers tend to prefer compositions with smaller $w_p$. In the bottom plot of Fig.~\ref{fig:FullComposition:3vs4moments:Cumulative:EPOS}, showing results in the highest considered energy bin, the fraction of heavy elements is at least $\sim20\%$ at $2\sigma$ level, with the best fit around $60\%$ and the upper limit consistent with no proton induced showers altogether.

\begin{figure}[!t]
    \centering
\includegraphics[width=0.8\linewidth]{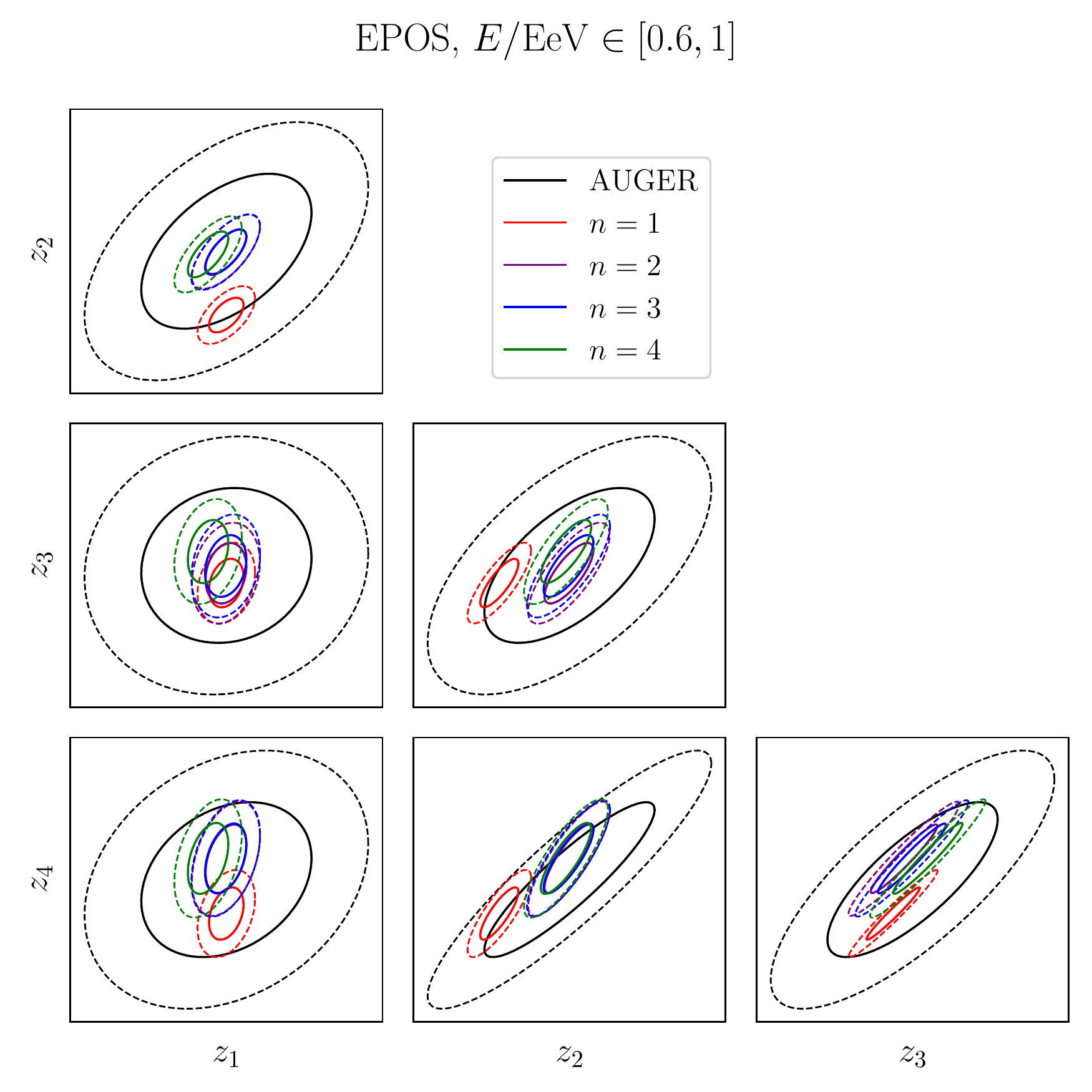}
\caption{$3\sigma$ (solid) and $5\sigma$ (dashed) contours of the multivariate moment distributions for the Auger data (black) and for the best composition inferred with EPOS, using $n=1$ (red), $n=2$ (purple), $n=3$ (blue) and $n=4$ (blue) moments, in the low energy bin, for the full $Z=26$ primary framework.}
\label{fig:FullComposition:3vs4moments:LowBin:z1-z4:EPOS}
\end{figure}

Next we note that results of fits to $n=3 (4)$ moments are again perfectly comparable, especially in terms of the inferred confidence intervals. Thus, at currently available statistics, three central moments suffice to extract the most relevant information on the composition even in the full $Z=26$ case. To further quantify the possible differences between results obtained with fits to different numbers of central moments, 
we show in Fig.~\ref{fig:FullComposition:3vs4moments:LowBin:z1-z4:EPOS} the $3\sigma$ (solid) and $5\sigma$ (dashed) contours of the multivariate normal distributions of moments in the low energy bin $E\in[0.6,~1]$~EeV, projected onto planes spanned by pairs of moments, see eq.~\eqref{eq:Moments:Multivariate} and \eqref{eq:Moments:Multivariate:Auger}. As in Fig.~\ref{fig:4Composition:3vs4moments:LowBin:z1-z4:EPOS}, the black ellipses indicate the contours of the measured Auger Open Data moment distribution, while the red, purple, blue and green lines indicate the maximum likelihood compositions, inferred from unbinned fits to $n=1,2,3,4$ central moments, respectively. Note that due to more statistics in this energy bin and larger composition space, when fitting to only $z_1 = \langle \Xm \rangle$, the best fit composition now clearly gives poor predictions for higher moments (inconsistent with the data at the $3\sigma$ level). This indicates that the additional information provided by higher central moments is relevant especially when trying to infer UHECR composition including more primaries from high enough statistics datasets. Currently including the second (and third) moments in the fit, the resulting predictions for the higher (third (and fourth)) moments are still consistent within uncertainties with the data, and with model results obtained when these higher moments are included in the fit. We expect the importance of higher moments to further increase when fitting to larger UHECR shower datasets already collected at the Pierre Auger Observatory. For completeness, we show the results for the intermediate and high energy bins in Figs.~\ref{fig:FullComposition:3vs4moments:IntermBin:z1-z4:EPOS} and \ref{fig:FullComposition:3vs4moments:HighBin:z1-z4:EPOS} respectively.

\subsection{Comparison of Hadronic Models}
\label{subsec:HMComaprison:Results}

\begin{figure}[!t]
\centering
\includegraphics[width=1\linewidth]{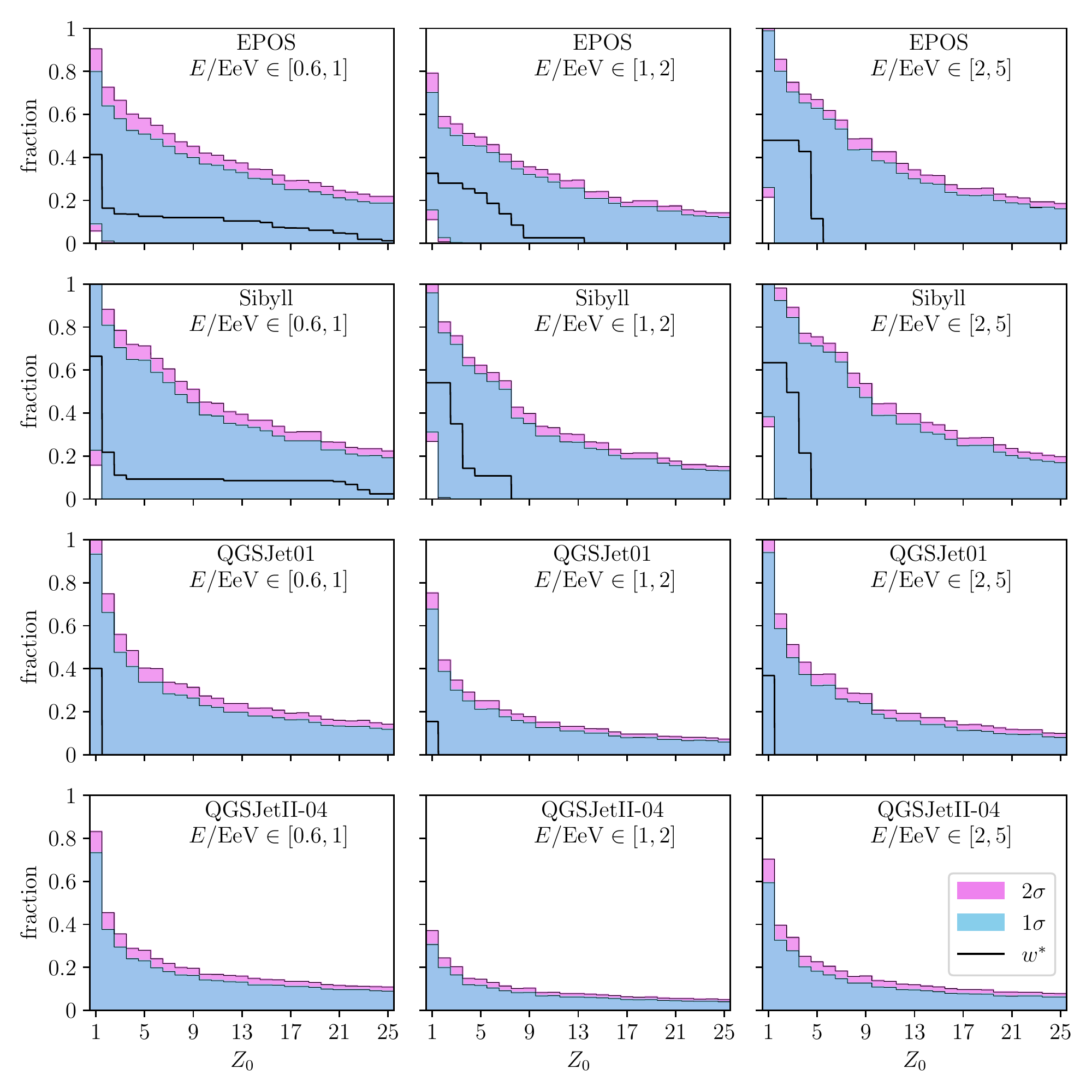}
\caption{Fraction of primaries with atomic number $Z > Z_0$ inferred with all four hadronic models and $n=3$, in the three energy bins considered. The black line shows the fraction for the best composition, while the cyan and purple regions indicate the $1\sigma$ and $2\sigma$ confidence levels respectively.
}\label{fig:FullComposition:4moments:Cumulative:AllModels}
\end{figure}

Finally we compare the results for the full UHECR composition based on simulations obtained with different hadronic models.
In Fig.~\ref{fig:FullComposition:4moments:Cumulative:AllModels} we show the  results for all four hadronic models considered in Sec.~\ref{subsec:Simulation}. It is immediately clear how these models can lead to very different conclusions. The Sibyll model (second row) tends to predict heavier compositions and smaller proton fractions than EPOS, with a lower limit of $\sim40\%$ at $2\sigma$ for the fraction of elements beyond  protons. On the other hand, the two QGSJet models (last two rows) favor light compositions, with all predictions consistent at $2\sigma$ with a 100\% proton shower composition.

\begin{table}[!t]
\begingroup
\setlength{\tabcolsep}{7pt} 
\renewcommand{\arraystretch}{1.3} 
\centering
\begin{tabular}{l|c c c} \hline \hline
                &  $E/{ \rm EeV}\in[0.6,1]$ &  $E/{ \rm EeV}\in[1,2]$ &  $E/{ \rm EeV}\in[2,5]$ \\  \hline 

    EPOS        & $2.1 ~(8.6) ~[19.5] ~\{37.0\}$ & $1.8 ~(8.1) ~[19.2] ~\{35.1\}$ & $1.9 ~(8.5) ~[19.9] ~\{36.4\}$ \\
    Sibyll 2.3c      & $2.1 ~(8.6) ~[19.5] ~\{38.2\}$ & $1.8 ~(8.0) ~[19.3] ~\{35.2\}$ & $1.9 ~(8.4) ~[19.8] ~\{35.9\}$ \\
    QGSJet01    & $2.1 ~(10.3) ~[23.1] ~\{44.4\}$ & $1.8 ~(10.7) ~[23.6] ~\{39.8\}$ & $2.0 ~(9.4) ~[22.3] ~\{38.3\}$ \\
    QGSJetII-04 & $3.9 ~(19.8) ~[31.2] ~\{50.3\}$ & $4.0 ~(19.4) ~[41.6] ~\{57.6\}$ & $3.5 ~(22.3) ~[41.8] ~\{59.6\}$ \\
    \hline \hline
    \end{tabular}
    \caption{Values of $-\log{\cal L}$ for the most probable $26$ dimensional compositions, for each hadronic model in three energy bins, using $n=1$ ($n=2$) [$n=3$] $\{n=4\}$ moments. Smaller values indicate a better fit to the measured data.}
    \label{tab:loglikelihood}
\endgroup
\end{table}

Another comparison of the four hadronic models is provided by the quality of their fits to Auger data. In Table~\ref{tab:loglikelihood} we summarize the values taken by the negative log-likelihood, Eq.~\eqref{eq:LogLikelihood}, at the best composition point in each energy bin, when considering different numbers of moments $n$. For the same $n$, smaller values indicate better fits to the measured Auger data in a given energy bin. With only a single feature, namely $\langle \Xm \rangle$, all models give similar results, that is all models fit the data equally well (but can infer markedly different compositions). Differences in the goodness of fit start to emerge only when increasing the number of  higher moments considered. We see that, both in the case of $n=3$ and $n=4$, the EPOS and Sibyll models provide the best fits, while the two QGSJet models yield significantly higher values of $-\log{\cal L}$.

\begin{figure}[!t]
\centering
        \includegraphics[width=.8\linewidth]{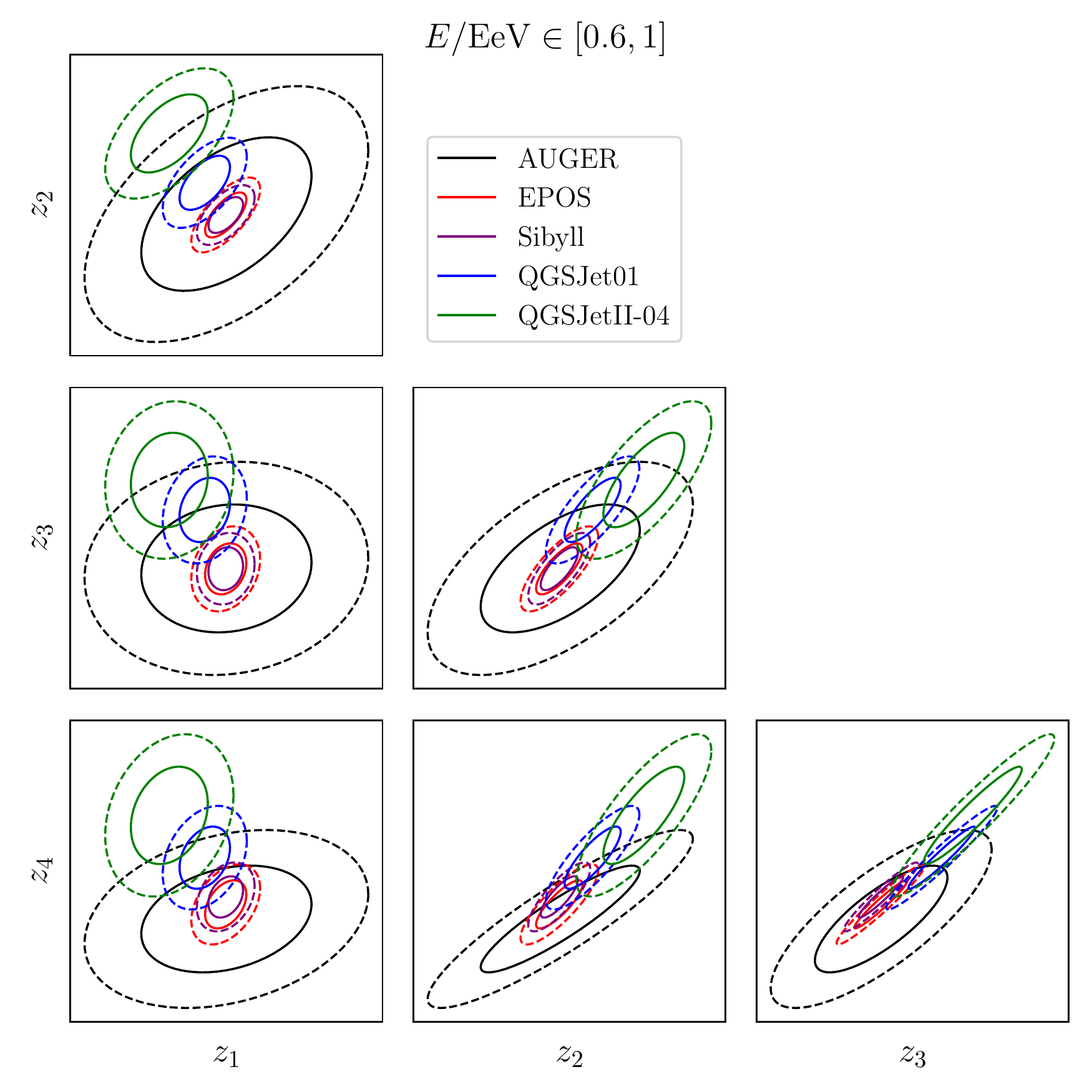}
        \caption{$3\sigma$ (solid) and $5\sigma$ (dashed) contours of the multivariate moment distributions for the Auger data (black) and for the best $n=3$ composition inferred with Sibyll (red), EPOS (purple), QGSJet01 (blue) and QGSJetII-04 (green) respectively, in the low energy bin, $E\in[0.65,1]$ EeV.}
        \label{fig:AllModels:LowBin:z1-z4}
\end{figure}

A more detailed understanding of the differences between models can also be obtained by plotting the moments of best fitted compositions for each model against the Auger data. This is shown in Fig.~\ref{fig:AllModels:LowBin:z1-z4} for the lowest energy bin, where the $3\sigma$ and $5\sigma$ contours of the multivariate normal distributions of moments, projected onto planes spanned by pairs of moments are shown, see eqs.~\eqref{eq:Moments:Multivariate:Auger} and eqs.~\eqref{eq:Moments:Multivariate}. The black ellipses indicate the contours of the Auger moment distribution, while the red, purple, blue and green lines are for the best fit Sibyill, EPOS, QGSJet01 and QGSJetII-04 models respectively. While the first two models sit inside the $3\sigma$ region for all moments, the last two clearly give a poor fit to Auger results, especially for higher moments. The same results for the other two energy bins are shown in Figs.~\ref{fig:AllModels:IntermBin:z1-z4} and \ref{fig:AllModels:HighBin:z1-z4}.

\section{Conclusions}
\label{sec:Conclusions}

We proposed a novel approach to the problem of inferring the nuclear composition of UHECRs from the measurements of fluorescent light spectra, the s.c. longitudinal profiles. We applied it to the data released in the Auger Open Data Set, which contains $\sim10\%$ of the total recorded showers. 

The position of the peak of the longitudinal profile, called $\Xm$, signals the maximum energy deposited from the shower in the atmosphere in the form of electromagnetic radiation. The $\Xm$ of a single shower can be related to both the initial energy and the atomic number $A$ of the primary particle. However the inherent stochastic nature of the showering process introduces large fluctuations. Inferring the primary nucleus of any single shower is thus at present intractable. On the other hand, the distribution of $\Xm$ in a selected energy bin can be used to infer on the composition of UHECRs in that energy region. 

Starting from this observation we introduced central moments of the $\Xm$ distributions as discriminating features of their primary components. To extract the composition from data, one has to rely on simulations, which in turn depend on the hadronic model assumed to compute the first series of interactions in the atmosphere. We performed our simulations with CORSIKA, using all four hadronic models available, in order to provide a quantitative comparison of their ability to fit the data and highlight their differences.

In our approach the distributions of moments of $\Xm$ as measured by Auger, are fit to the distributions of moments of simulated $\Xm$. The $\Xm$ are simulated for each single primary with $Z = 1,\dots,26$, that is from proton to iron, and then combined assuming different compositions. A number of convenient simplifications and approximations allows for the likelihood to be expressed in a compact form and computed efficiently for any assumed composition. Finally, the computationally intensive task of covering the high-dimensional space of all possible compositions is tackled using Nested Sampling algorithms to estimate the likelihoods of the compositions and their confidence regions.

Our method differs from existing approaches in the literature in several significant ways. Firstly, owing to the implementation of the nested sampling technique and efficient likelihood evaluation, we were able to explore for the first time the full range of possible compositions, while previous works limited themselves to mixtures of only a few nuclei. In addition, all previous analyses used the binned $\Xm$ distributions directly to fit the compositions. The main discriminating features that  differentiate between different compositions and/or hadronic models however remained obscured. In addition, in the binned likelihood approach, it is often difficult to discern the effects of systematic  and statistical uncertainties (of measurements as well as simulations) on the final results. Our unbinned likelihood approach, based on the systematic characterisation of $\Xm$ distributions in terms of their first few central moments, addresses both of these issues. In particular, it allows to transparently include systematic and statistical uncertainties in the fit, both from the data and Monte-Carlo simulations. In addition the discrimination power of individual moments is easily identified allowing for transparent model and composition comparison.

Finally, since the central moments of $\Xm$ distributions, conveying their most relevant features, can be systematically and efficiently computed, they are suitable for further studies and improvements. In particular, larger statistics datasets available to the Auger collaboration could potentially warrant the inclusion of higher moments in the fits. Certainly they should more strongly constrain the allowed compositions of UHECRs and allow to better discriminate between different hadronic models. Potentially they could even allow to probe the presence of exotic primaries (such as leptons or new hypothetical massive (quasi)stable particles). In addition, the compact form of the likelihood and transparency of the main discriminating features in our approach should facilitate the application of Machine Learning methods in the analysis of UHECRs. We leave all of these explorations for future work.

\section*{Acknowledgments}
The authors would like to thank Vid Hom\v sak for his involvement in the initial stages of the project. JFK deeply appreciates the continuing hospitality and computing support of CERN without which this project would not have been possible.  The authors acknowledge the financial support from the Slovenian Research Agency (grant No. J1-3013 and research core funding No. P1-0035).


\begin{appendix}
%
\section{Additional plots}
\label{app:extraplots}
%

\begin{figure}[!h]
\centering
~~~~~~~~~~~~\includegraphics[width=0.8\linewidth]{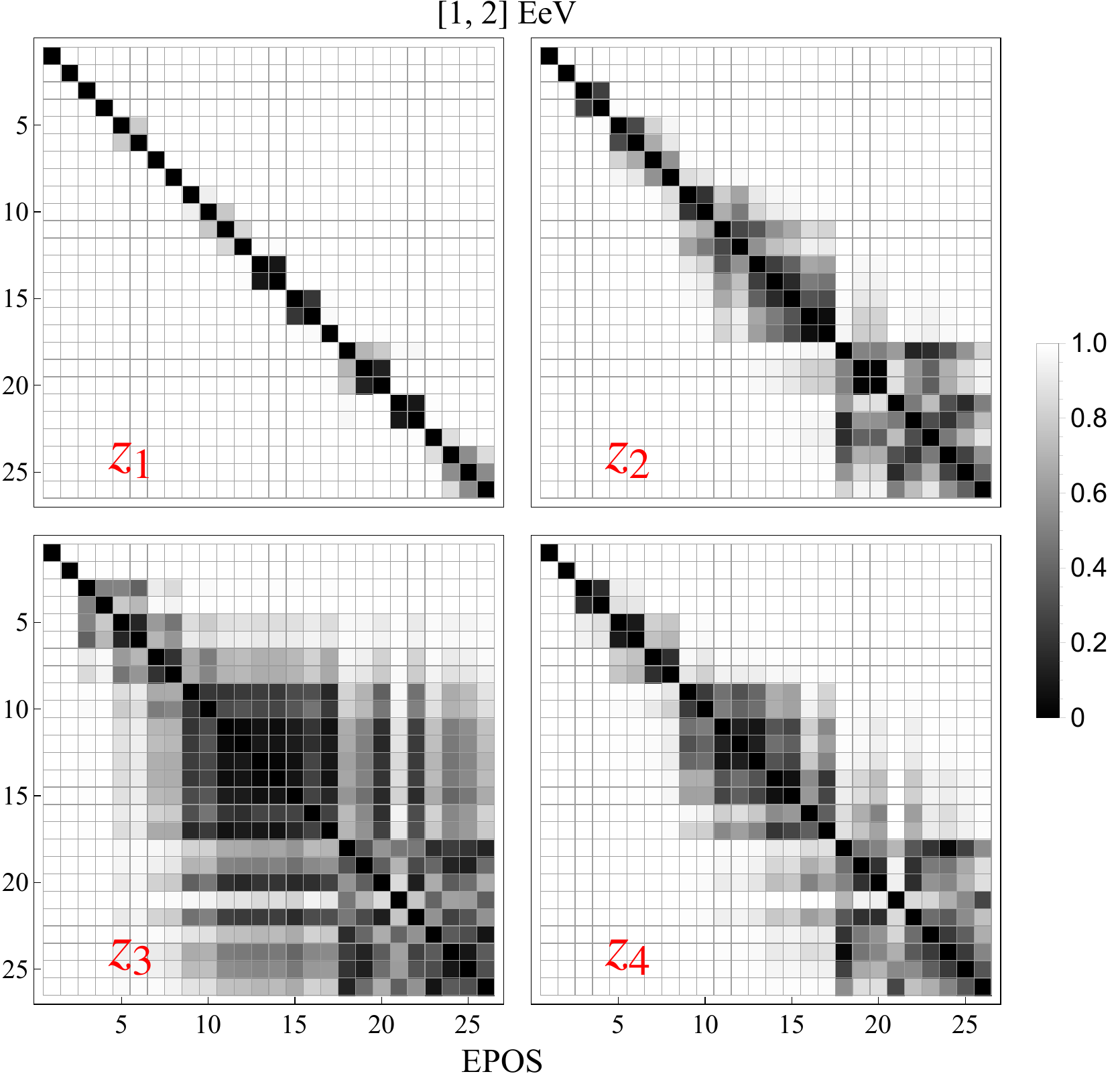}
\caption{Hellinger distance $\mathcal H_{ij}(z_n)$ between different primaries $i,j$ within the EPOS model for the first four central moments $z_n$ of $\Xm$ distributions of simulated UHECRs with energies within $[1,2]$~EeV. See text in Sec.~\ref{subsec:ModelComparison} for details.}\label{fig:Hellinger:EPOS}
\end{figure}

\begin{figure}[!h]
\centering
~~~~~~~~~~~~\includegraphics[width=0.8\linewidth]{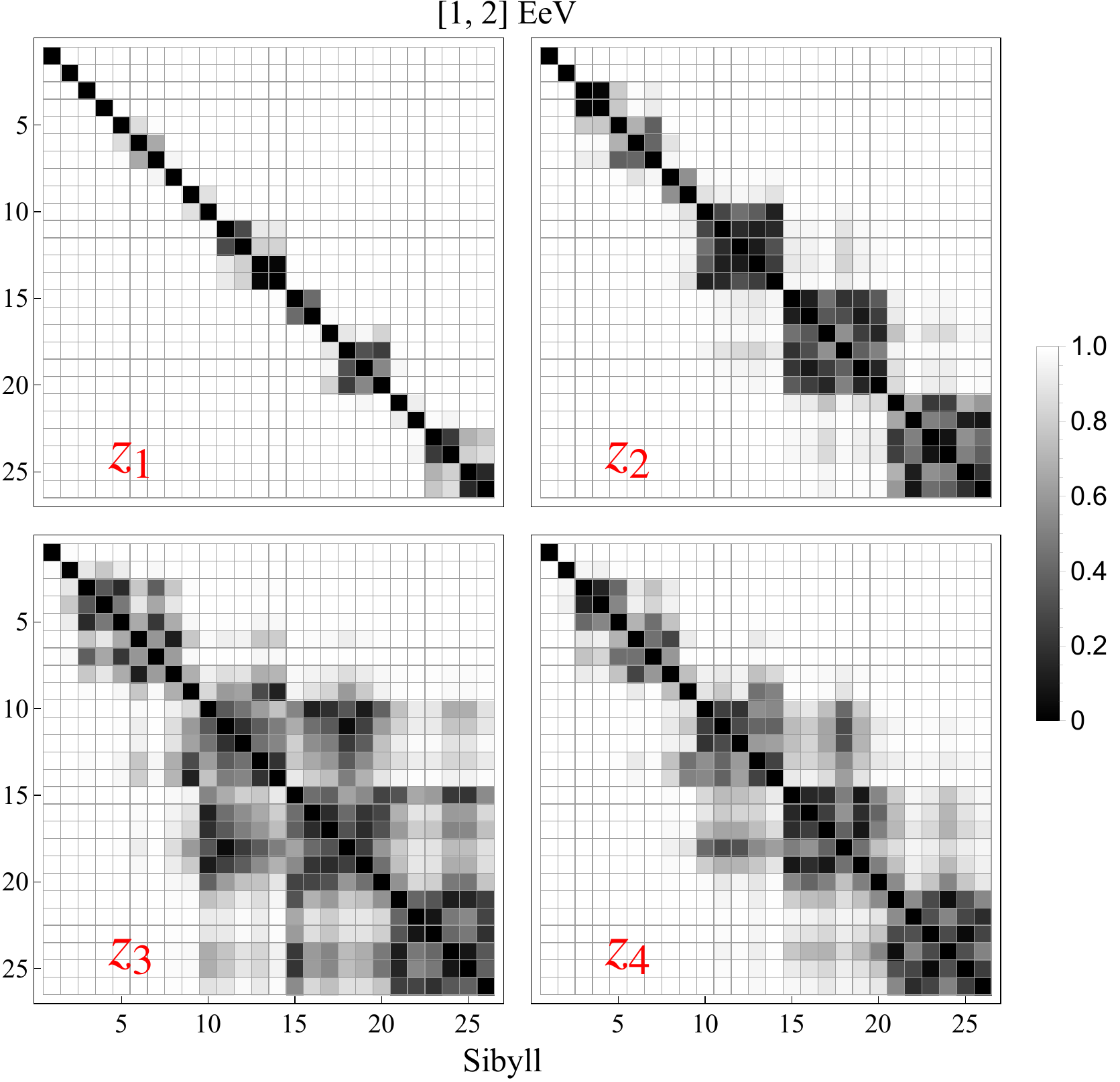}
\caption{Same as Fig.~\ref{fig:Hellinger:EPOS}, for Sibyll model.}\label{fig:Hellinger:SIB}
\end{figure}

\begin{figure}[!h]
\centering
~~~~~~~~~~~~\includegraphics[width=0.8\linewidth]{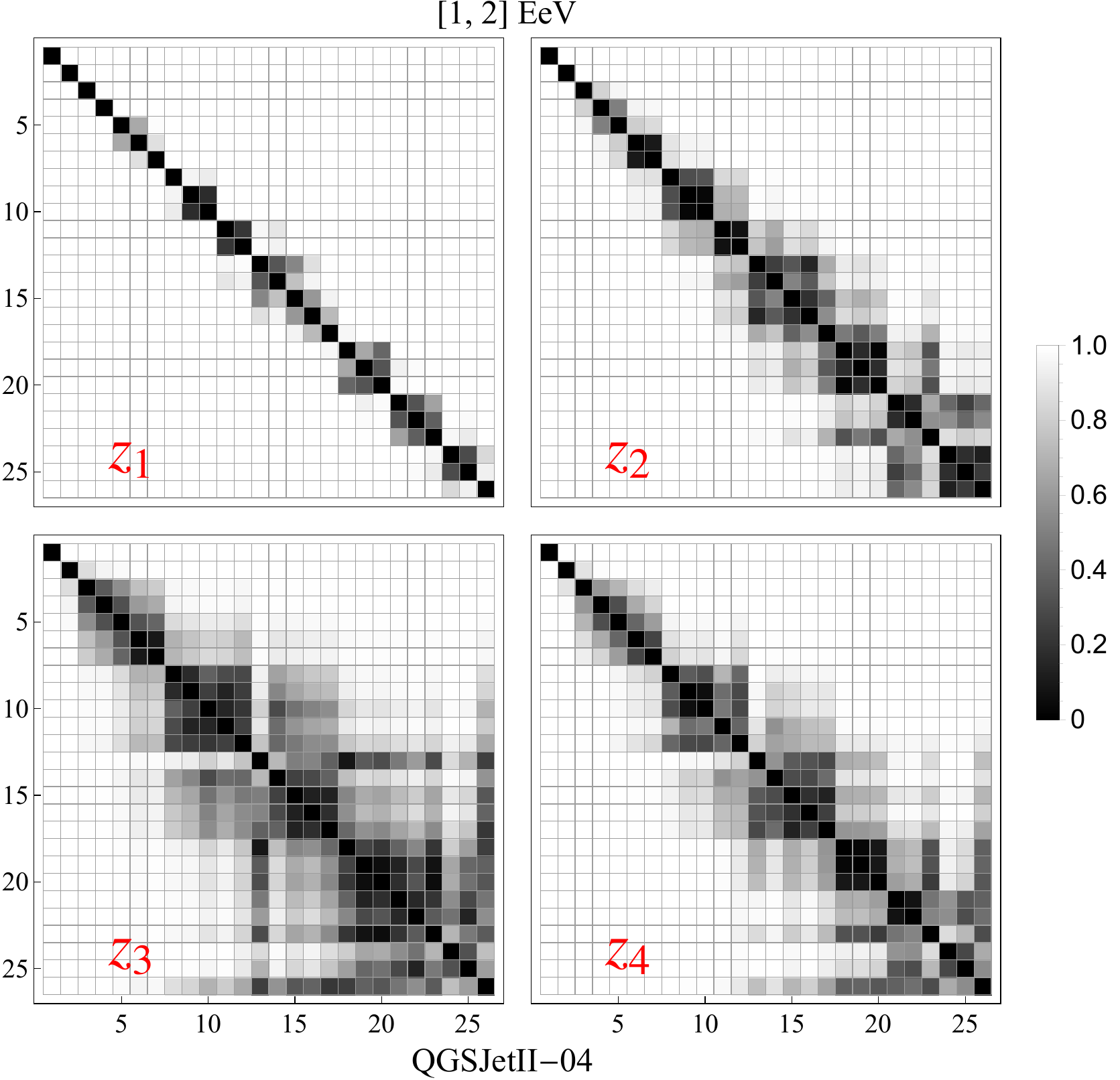}
\caption{Same as Fig.~\ref{fig:Hellinger:EPOS}, for QGSJetII-04 model.}\label{fig:Hellinger:QGS4}
\end{figure}

\begin{figure}[!h]
\centering
~~~~~~~~~~~~\includegraphics[width=0.8\linewidth]{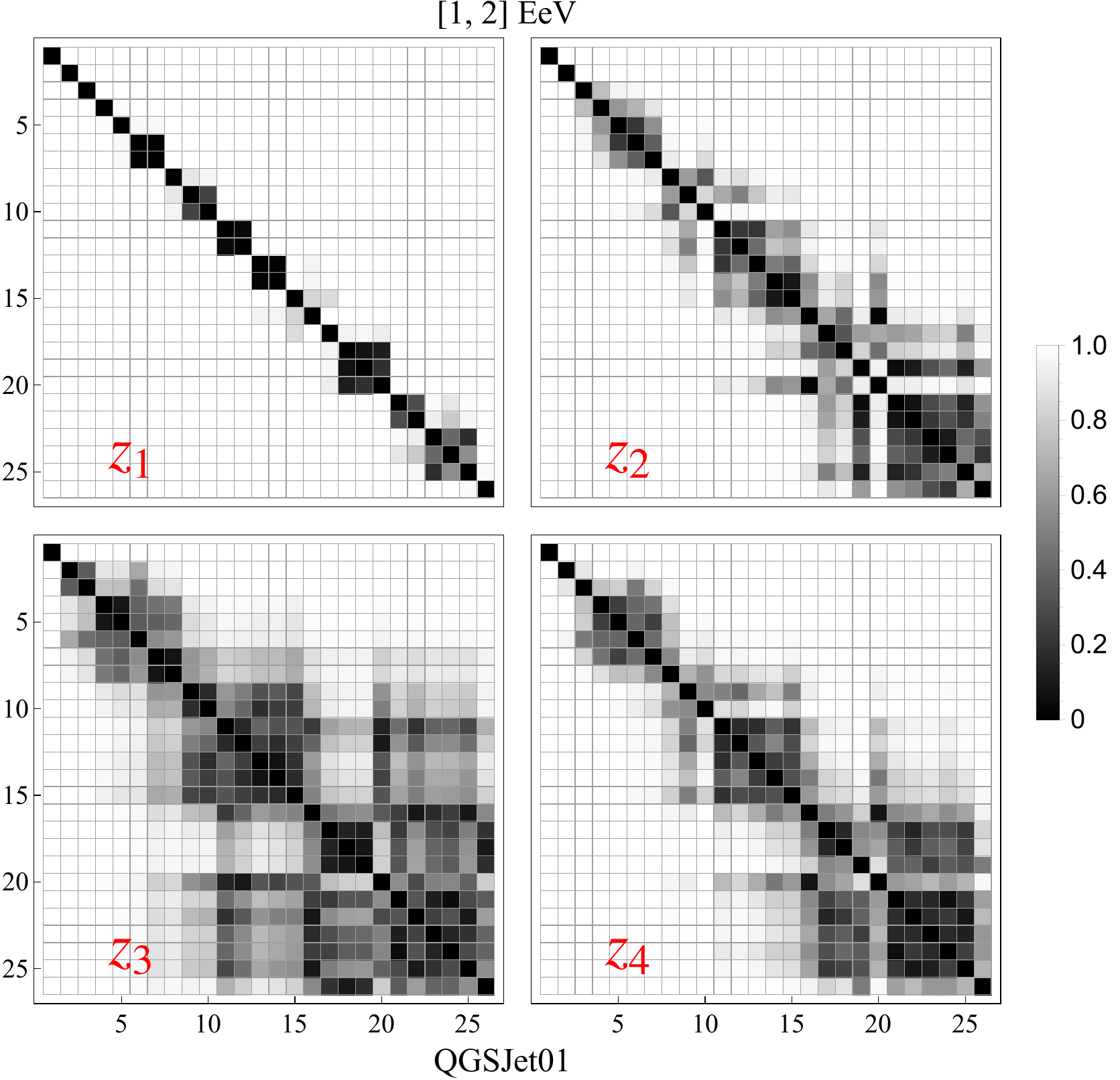}
\caption{Same as Fig.~\ref{fig:Hellinger:EPOS}, for QGSJet01 model.}\label{fig:Hellinger:QGS5}
\end{figure}

\begin{figure}[!h]
\centering
~~~~~~~~~~~~\includegraphics[width=0.8\linewidth]{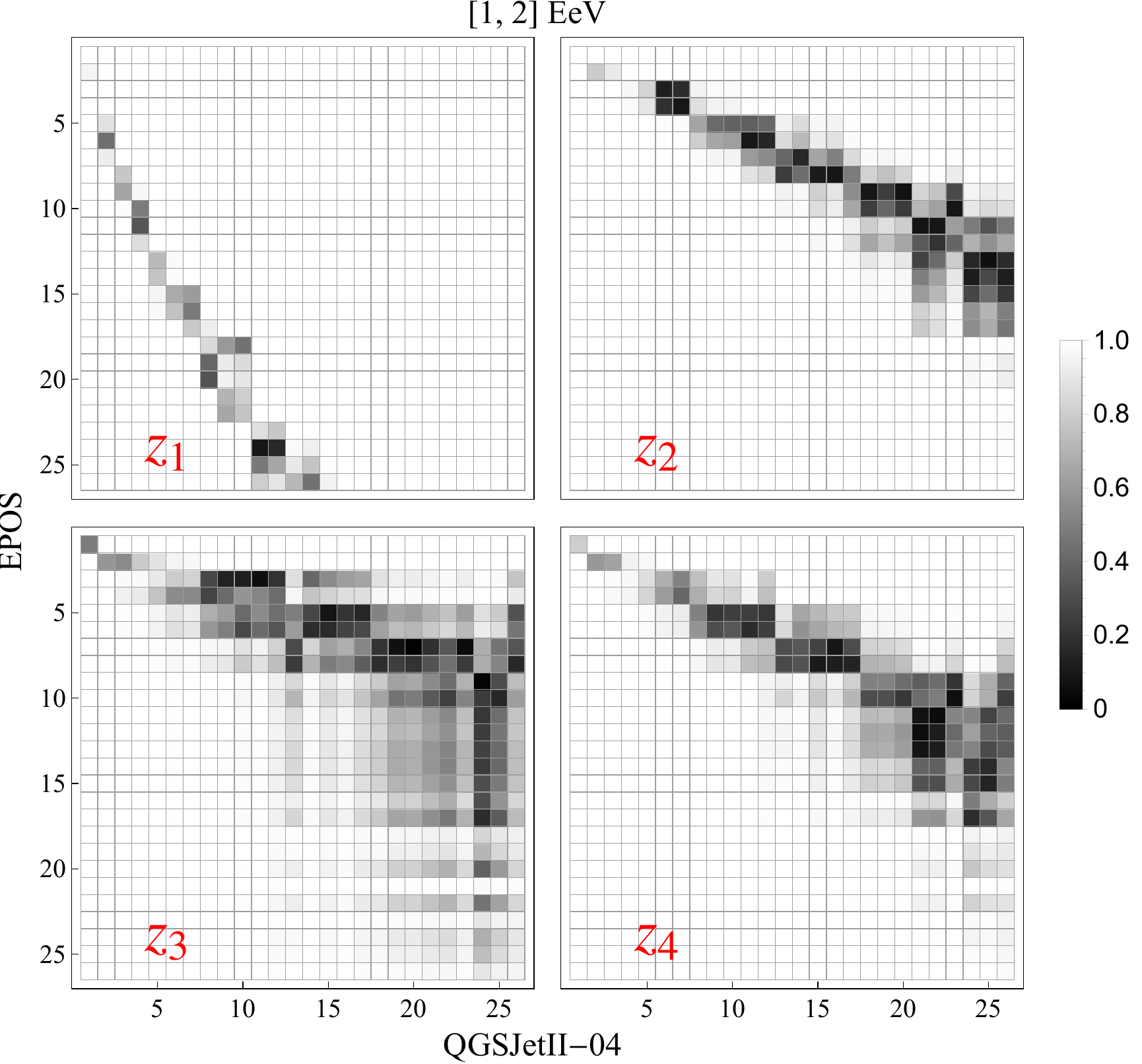}
\caption{Same as Fig.~\ref{fig:EPOSvsSIB}, for EPOS and QGSJetII-04 models.}\label{fig:EPOSvsQGS4}
\end{figure}

\begin{figure}[!h]
\centering
~~~~~~~~~~~~\includegraphics[width=0.8\linewidth]{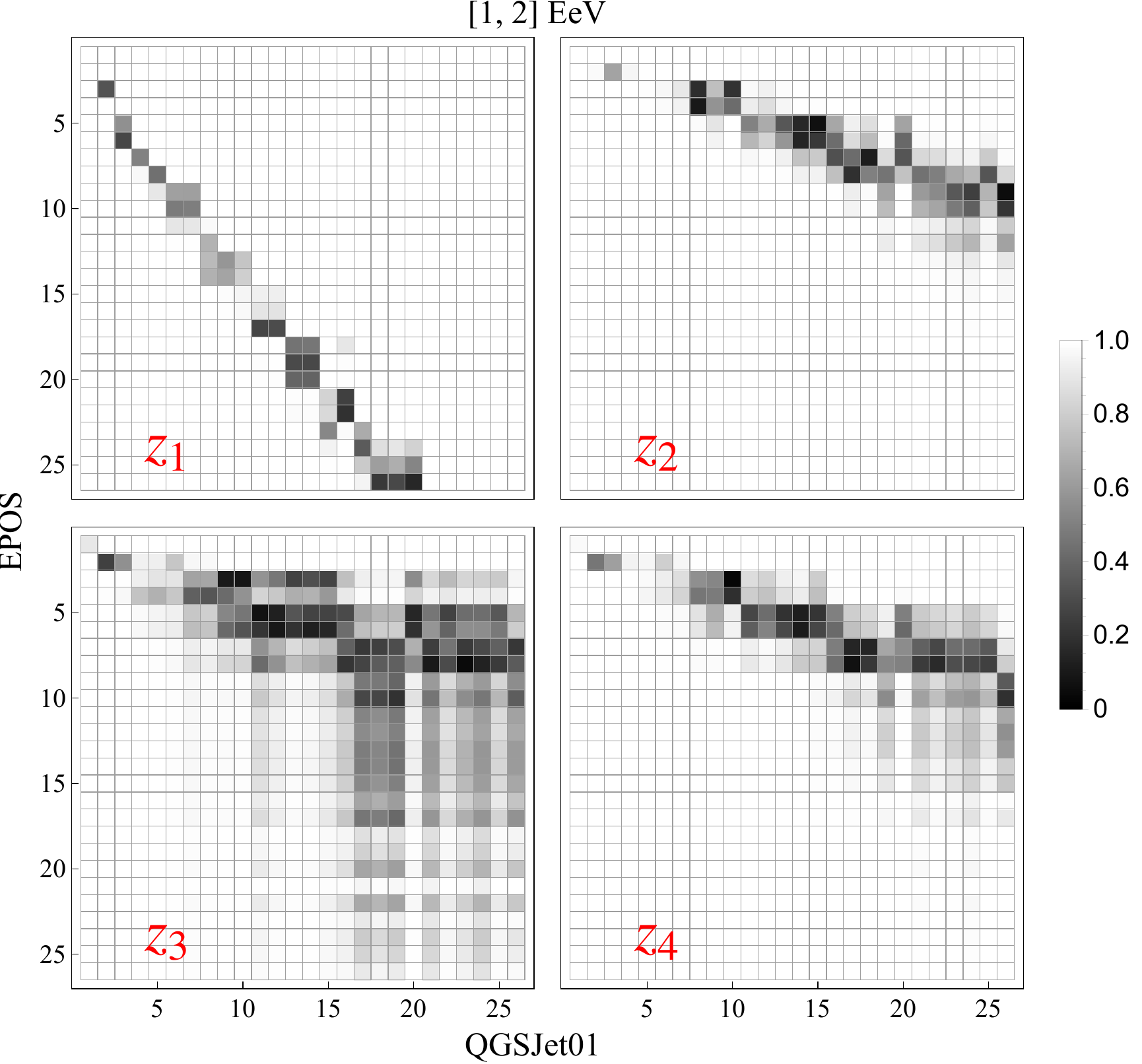}
\caption{Same as Fig.~\ref{fig:EPOSvsSIB}, for EPOS and QGSJet01 models.}\label{fig:EPOSvsQGS5}
\end{figure}

\begin{figure}[!h]
\centering
~~~~~~~~~~~~\includegraphics[width=0.8\linewidth]{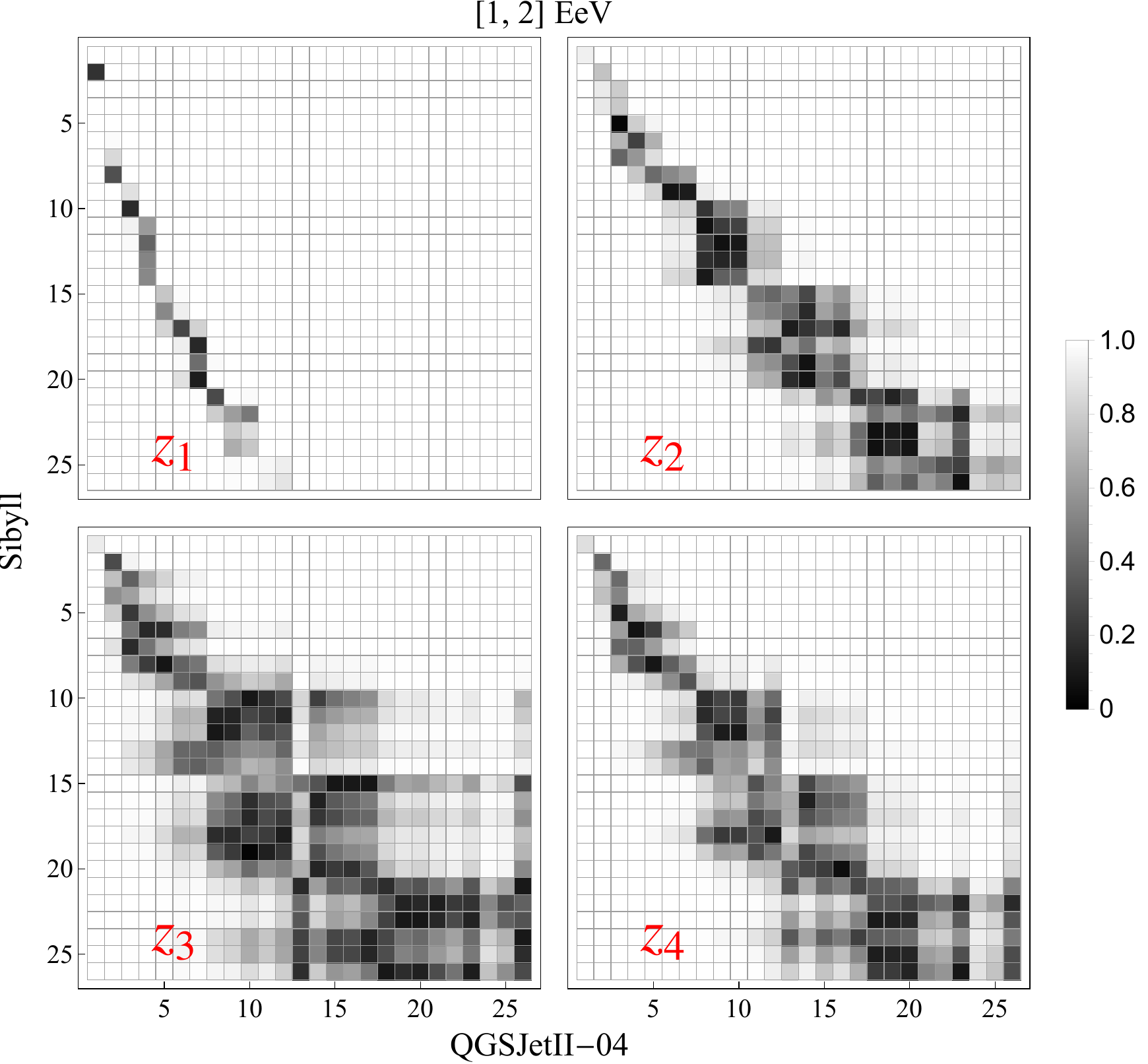}
\caption{Same as Fig.~\ref{fig:EPOSvsSIB}, for Sibyll and QGSJetII-04 models.}\label{fig:SIBvsQGS4}
\end{figure}

\begin{figure}[!h]
\centering
~~~~~~~~~~~~\includegraphics[width=0.8\linewidth]{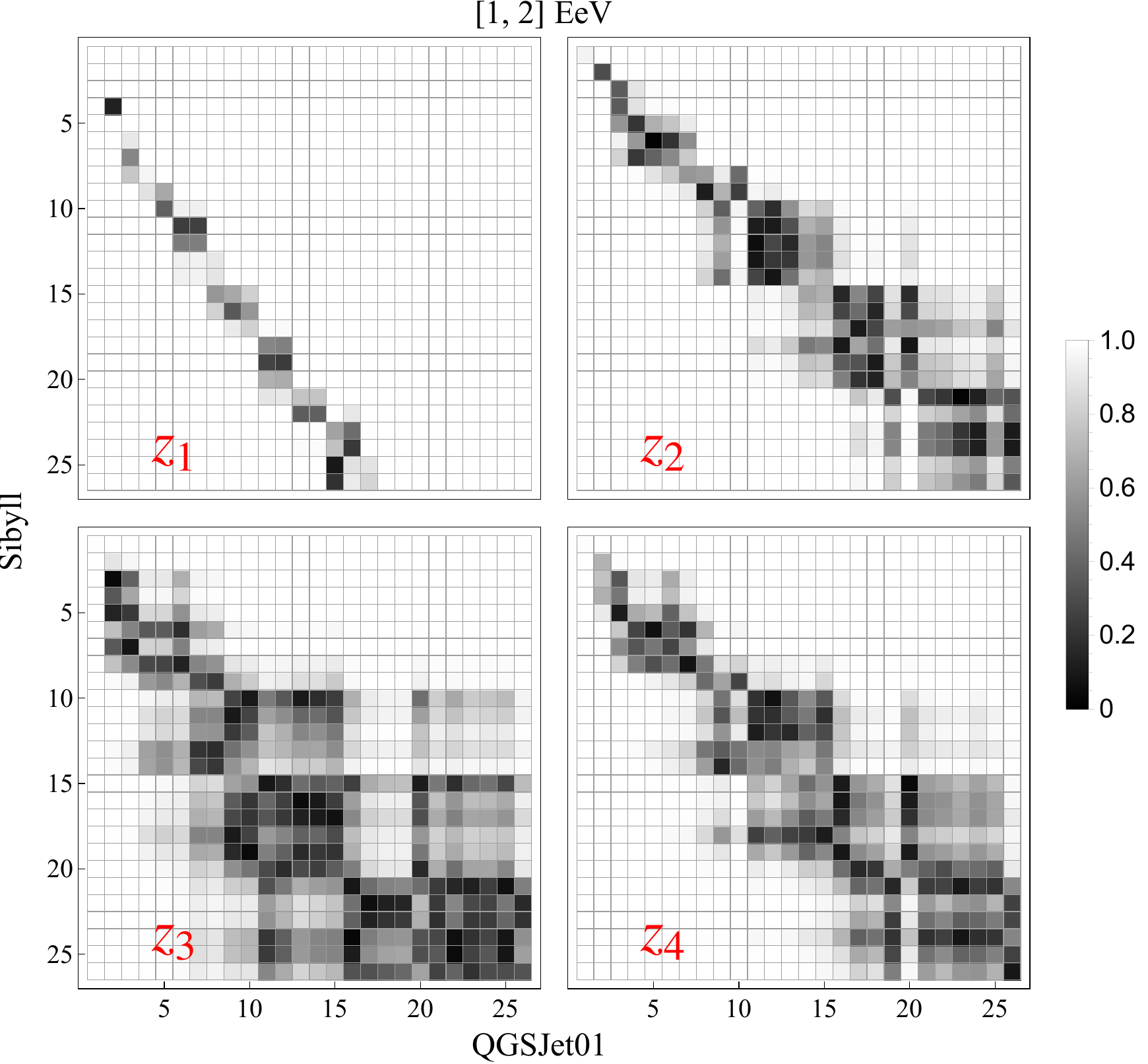}
\caption{Same as Fig.~\ref{fig:EPOSvsSIB}, for Sibyll and QGSJet01 models.}\label{fig:SIBvsQGS5}
\end{figure}

\begin{figure}[!h]
\centering
~~~~~~~~~~~~\includegraphics[width=0.8\linewidth]{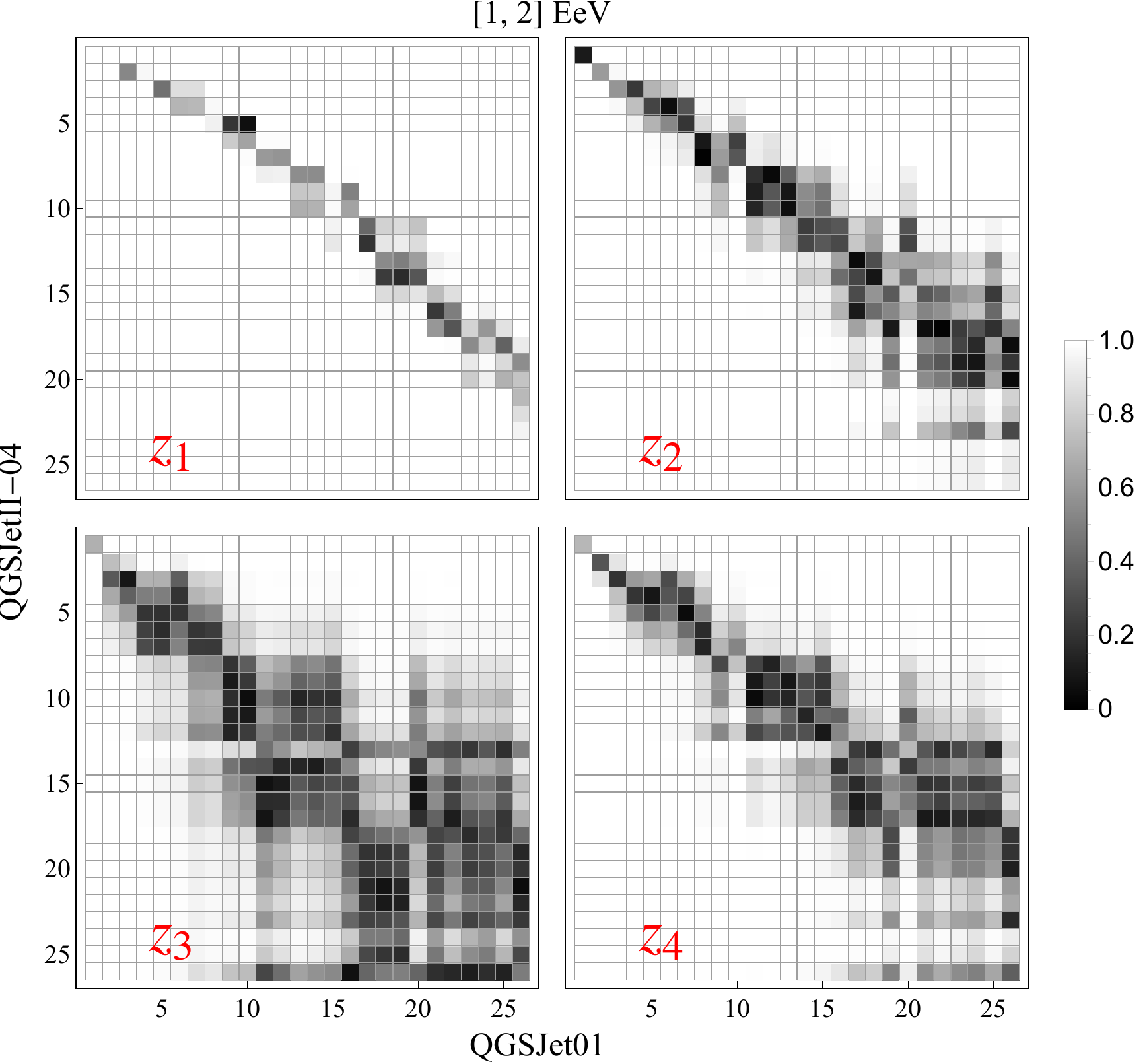}
\caption{Same as Fig.~\ref{fig:EPOSvsSIB}, for QGSJetII-04 and QGSJet01 models.}\label{fig:QGS4vsQGS5}
\end{figure}

Figs~\ref{fig:Hellinger:EPOS}, \ref{fig:Hellinger:SIB}, \ref{fig:Hellinger:QGS4} and \ref{fig:Hellinger:QGS5} show the Hellinger distance matrix for different primaries simulated in the $[1,2]$~EeV energy bin, within the EPOS, Sibyll, QGSJetII-04 and QGSJet01 models respectively. Comparison of model pairs in the same energy bin are shown in Figs.~\ref{fig:EPOSvsQGS4}, \ref{fig:EPOSvsQGS5}, \ref{fig:SIBvsQGS4}, \ref{fig:SIBvsQGS5} and \ref{fig:QGS4vsQGS5}. For details on these plots, see the discussion in Sec.~\ref{subsec:ModelComparison}.

\begin{figure}[!h]
     \centering
 \includegraphics[width=0.8\linewidth]{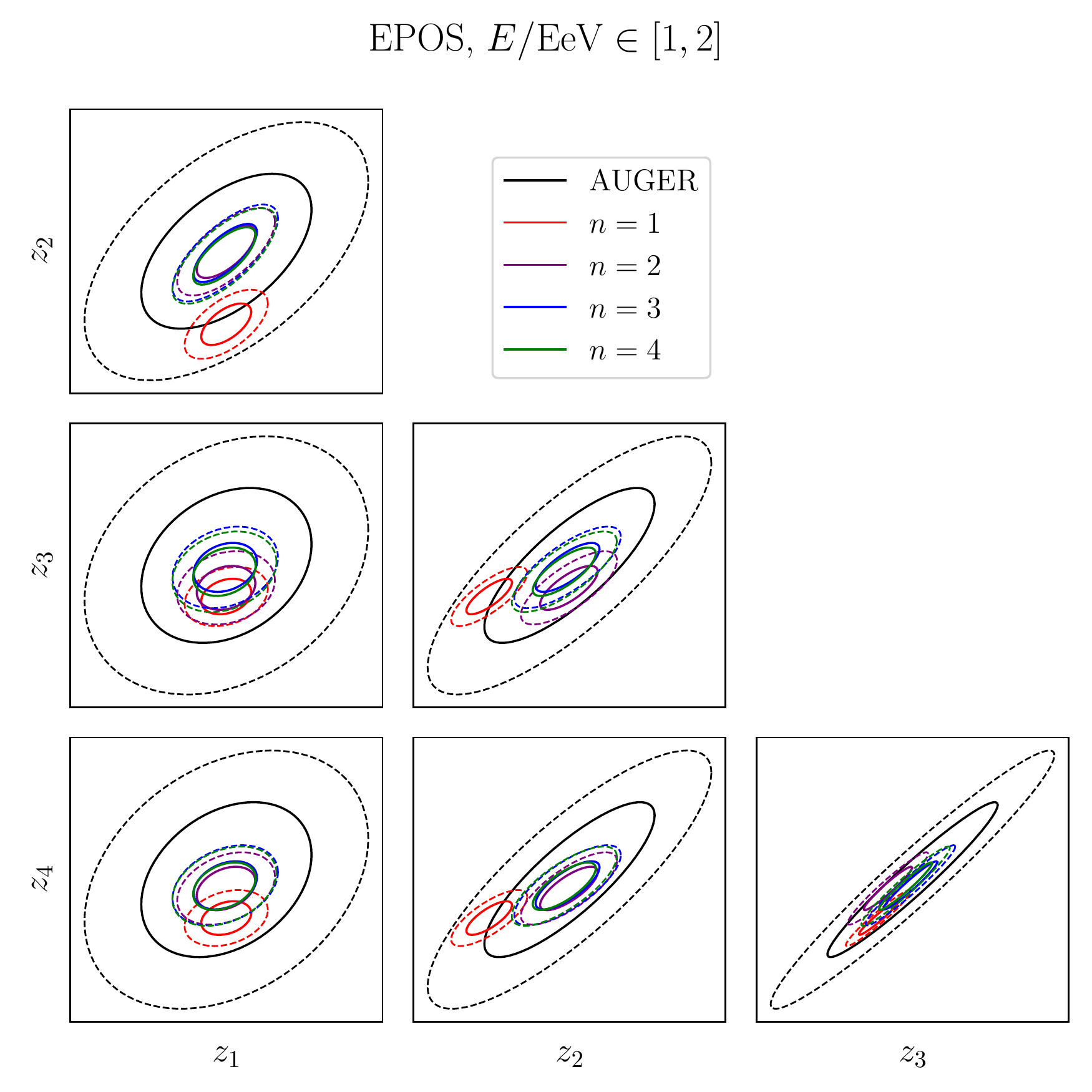}
 \caption{Same as Fig.~\ref{fig:FullComposition:3vs4moments:LowBin:z1-z4:EPOS} for the intermediate energy bin, $E\in[1,2]$ EeV.}\label{fig:FullComposition:3vs4moments:IntermBin:z1-z4:EPOS}
 \end{figure}

\begin{figure}[!h]
     \centering
 \includegraphics[width=0.8\linewidth]{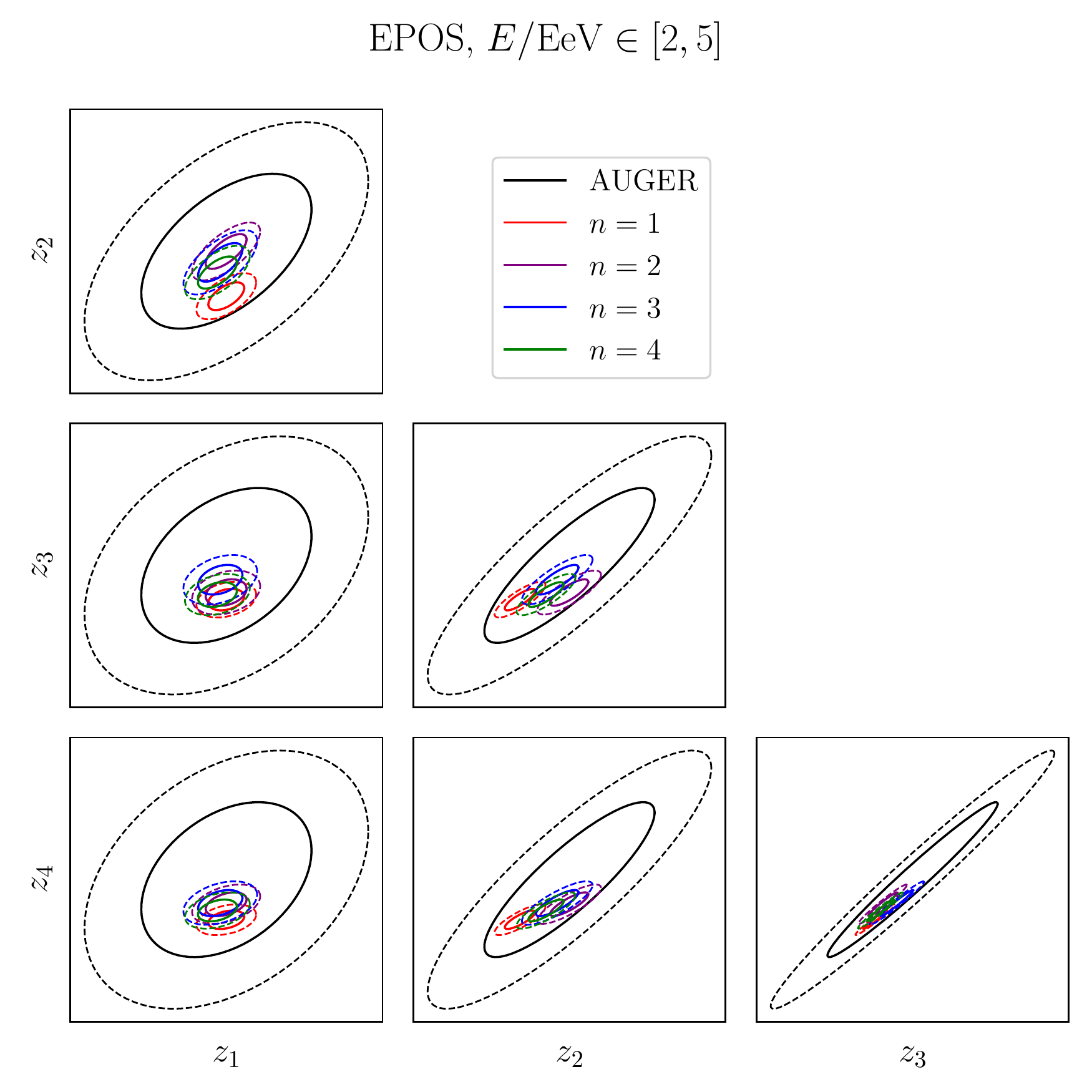}
 \caption{Same as Fig.~\ref{fig:FullComposition:3vs4moments:LowBin:z1-z4:EPOS} for the high energy bin, $E\in[2,5]$ EeV.}\label{fig:FullComposition:3vs4moments:HighBin:z1-z4:EPOS}
 \end{figure}

\begin{figure}[!h]
\centering
\includegraphics[width=0.8\linewidth]{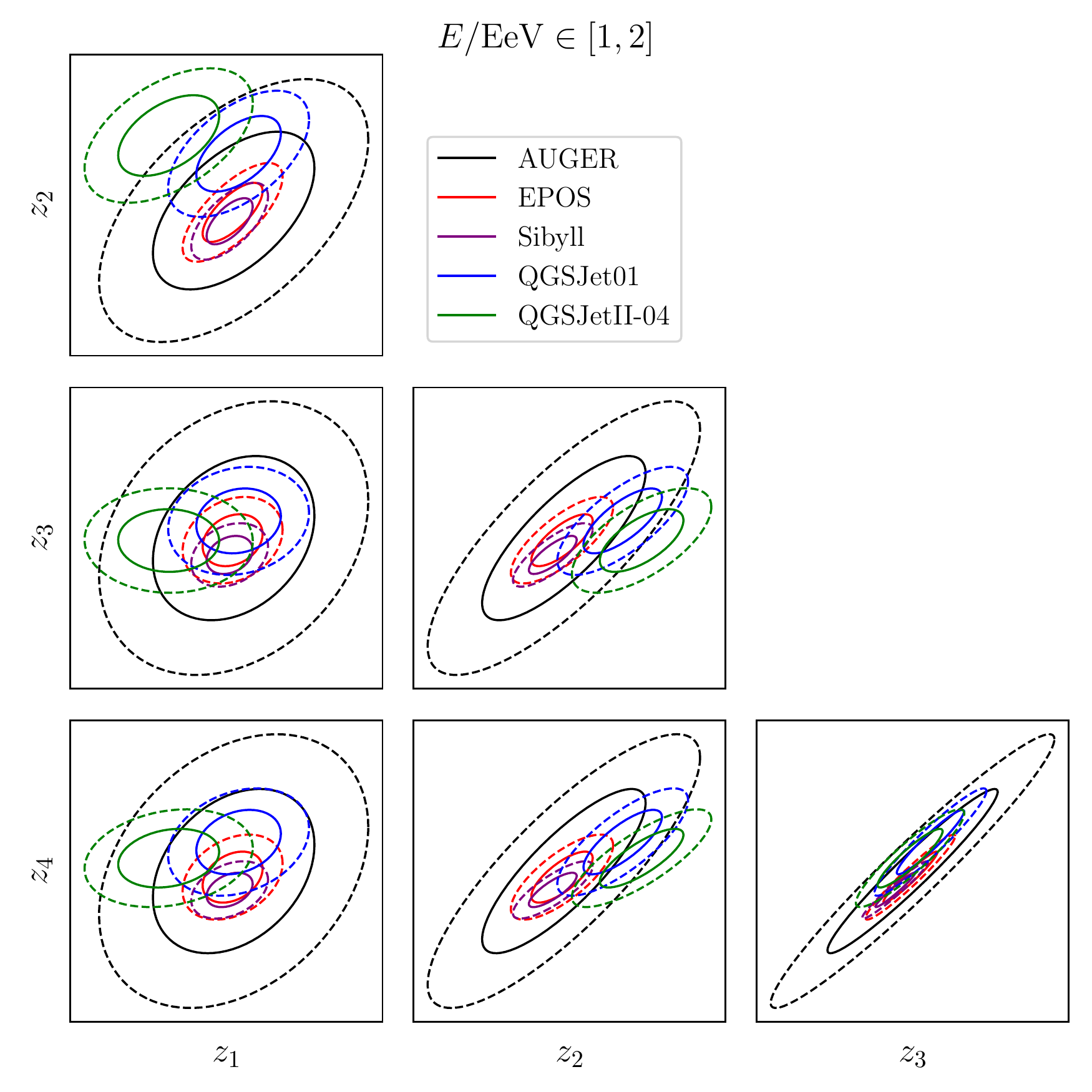}
\caption{Same as Fig.~\ref{fig:AllModels:LowBin:z1-z4}, for the intermediate energy bin, $E\in[1,2]$ EeV. }\label{fig:AllModels:IntermBin:z1-z4}
\end{figure}

\begin{figure}[!h]
\centering
\includegraphics[width=0.8\linewidth]{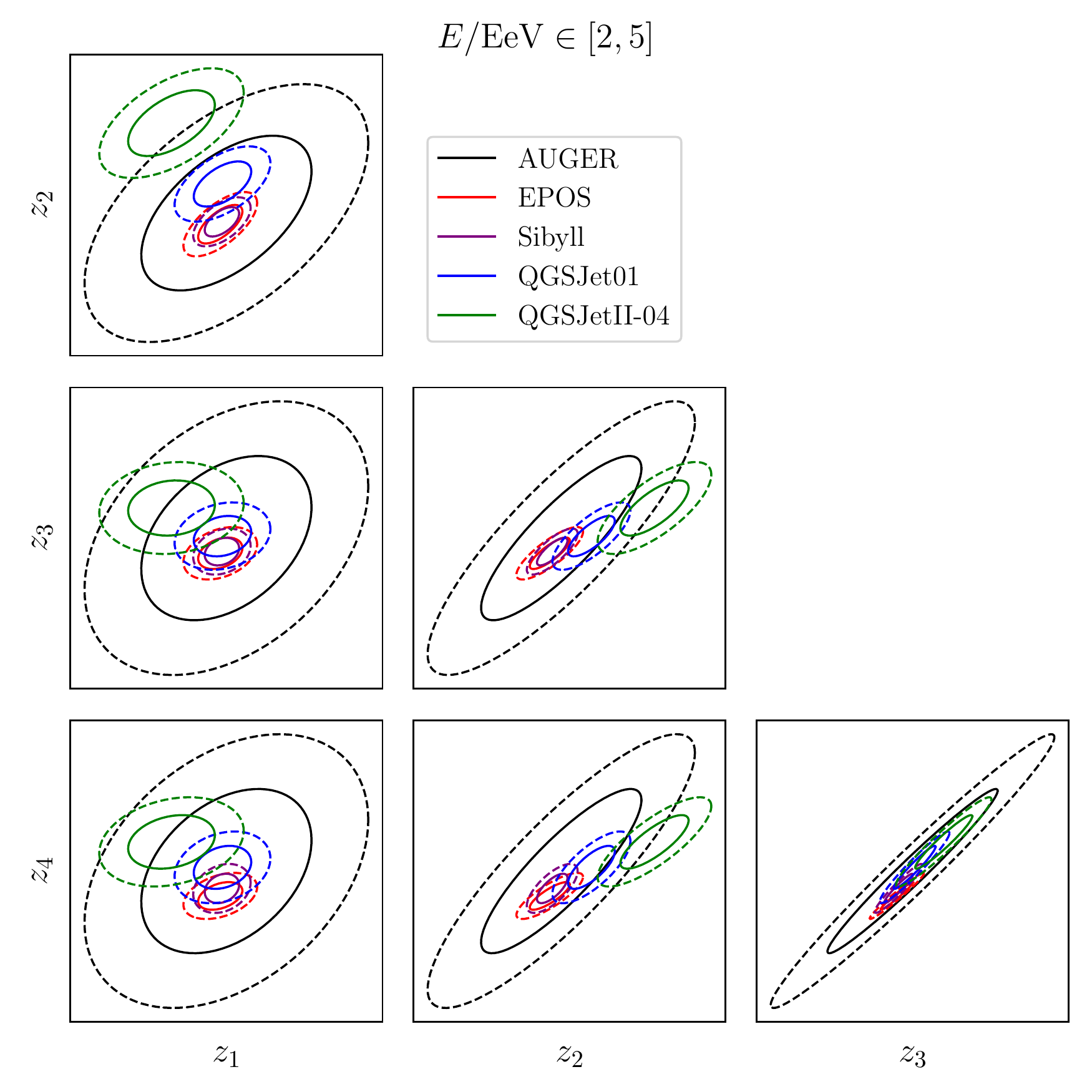}
\caption{Same as Fig.~\ref{fig:AllModels:LowBin:z1-z4}, for the high energy bin, $E\in[2,5]$ EeV. }\label{fig:AllModels:HighBin:z1-z4}
\end{figure}

Fig.~\ref{fig:FullComposition:4moments:Cumulative:AllModels} shows the results on the cumulative composition for all four hadronic models. Figs.~\ref{fig:AllModels:IntermBin:z1-z4} and \ref{fig:AllModels:HighBin:z1-z4} show the moment correlations of best compositions in the intermediate and high energy bin respectively. Finally, Figs.~\ref{fig:FullComposition:3vs4moments:IntermBin:z1-z4:EPOS} and \ref{fig:FullComposition:3vs4moments:HighBin:z1-z4:EPOS} show the comparison of the best compositions obtained with EPOS model, using $n=3$ or $n=4$ moments, in the intermediate and high energy bin respectively. The main discussion for the plots listed above can be found in Sec.~\ref{subsec:Fullcomposition:Results}.

\end{appendix}

\clearpage

\bibliographystyle{h-physrev}
\bibliography{references}

\end{document}